\pdfoutput=1
\documentclass[a4paper,fleqn,usenatbib]{mnras}

\usepackage[T1]{fontenc}
\usepackage{ae,aecompl}

\usepackage{graphicx}	
\usepackage{amsmath}	
\usepackage{amssymb}	

\title[Too small to succeed]{Too small to succeed: the difficulty of sustaining star formation in low-mass haloes}

\author[Cashmore, Wilkinson, Power \& Bourne]{
Claire R. Cashmore,$^{1}$\thanks{E-mail: crc16@le.ac.uk}
Mark I. Wilkinson,$^{1}$
Chris Power$^{2}$
and Martin A. Bourne$^{1,3}$
\\
$^{1}$ Department of Physics \& Astronomy, University of Leicester, University Road, Leicester LE1 7RH \\
$^{2}$ International Centre for Radio Astronomy Research, University of Western Australia, 35 Stirling Highway, Crawley, Western Australia 6009, Australia \\
$^{3}$ Institute of Astronomy and Kavli Institute for Cosmology, University of Cambridge, Madingley Road, Cambridge CB3 0HA, UK
}

\date{Accepted XXX. Received YYY; in original form ZZZ}

\pubyear{2017}

\begin{document}
\label{firstpage}
\pagerange{\pageref{firstpage}--\pageref{lastpage}}
\maketitle

\begin{abstract}

We present high resolution simulations of an isolated dwarf spheroidal (dSph) galaxy between redshifts $z\sim10$ and $z\sim 4$, the epoch when several Milky Way dSph satellites experienced extended star formation, in order to understand in detail the physical processes which affect a low-mass halo?s ability to retain gas. It is well-established that supernova feedback is very effective at expelling gas from a $3\times 10^7$M$_\odot$ halo, the mass of a typical redshift 10 progenitor of a redshift 0 halo with mass $\sim10^9$M$_\odot$. We investigate the conditions under which such a halo is able to retain sufficient high-density gas to support extended star formation. In particular, we explore the effects of: an increased relative concentration of the gas compared to the dark matter; a higher concentration dark matter halo; significantly lower supernova rates; enhanced metal cooling due to enrichment from earlier supernovae. We show that disk-like gas distributions retain more gas than spherical ones, primarily due to the shorter gas cooling times in the disk. However, a significant reduction in the number of supernovae compared to that expected for a standard initial mass function is still needed to allow the retention of high density gas. We conclude that the progenitors of the observed dSphs would only have retained the gas required to sustain star formation if their mass, concentration and gas morphology were already unusual for those of a dSph-mass halo progenitor by a redshift of 10.
\end{abstract}

\begin{keywords}
Galaxies: dwarf -- galaxies: ISM -- galaxies: formation -- galaxies: evolution -- stars: supernova -- ISM: supernova
\end{keywords}


\begingroup
\let\clearpage\relax
\pdfoutput=1
\section{Introduction}

Our current galaxy formation paradigm, hierarchical structure formation in the context of a $\Lambda$CDM cosmology, has had many notable successes on Mpc scales~\citep[e.g.][]{2015arXiv150702704P,2015arXiv150201589P,1985ApJ...292..371D,1996ApJ...457L..51H,1998AJ....116.1009R,2007ApJS..170..377S}. The low-luminosity tail of the galaxy luminosity function remains problematic, however, with both the numbers and properties of observed stellar systems occupying dark matter haloes with masses below $10^9$M$_\odot$ apparently in tension with the predictions of cosmological simulations~\citep[e.g][]{1999ApJ...522...82K, 2011MNRAS.415L..40B,2012MNRAS.422.1203B, 2010MNRAS.408.2364S}. Given our incomplete understanding of the interplay between baryons and dark matter in the context of galaxy formation, many authors have suggested that baryon physics may be able to account for all the apparent discrepancies between $\Lambda$CDM and observations~\citep{1996MNRAS.283L..72N,2012MNRAS.421.3464P}.


It is widely expected that there should be a lower limit on the masses of dark matter haloes which host observable galaxies, of any luminosity, at $z=0$~\citep[e.g.][]{2006MNRAS.371..885R,2016MNRAS.tmp..504R,2002ApJ...572L..23S,2016MNRAS.456...85S}. To date, the resolution limits of numerical studies of low-mass galaxy formation in a cosmological context have meant that few meaningful results have been obtained for objects with stellar masses below $10^6$M$_{\odot}$, effectively leaving out the majority of the dwarf satellites of the Milky Way (MW)~\citep[see][for the highest resolutions studies to date]{2012ApJ...761...71Z,2012MNRAS.422.1231G,2015MNRAS.448.2941S,2015MNRAS.454.2092O,2015MNRAS.454.2039F}.  The lowest luminosity classical dwarf spheroidals (dSphs) such as Ursa Minor and Draco have stellar masses of a few\,$\times10^5$M$_{\odot}$~\citep[]{2012AJ....144....4M}, while the ultra faint dwarfs (UFDs) are still less luminous~\citep[]{2006ApJ...643L.103Z,2006ApJ...650L..41Z,2007ApJ...654..897B,2008ApJ...688..277T}. These pose particular challenges for numerical simulations as a complete physical picture necessitates following both internal and external processes on timescales which span at least 5 orders of magnitude. For example, the evolution of individual SNa remnants occurs on timescales of $10^4$ years while ram pressure stripping of gas due to the dSph's orbit around the Milky Way modulates its evolution on timescales of $10^8-10^9$ years. 

Abundance matching methods attempt to link the dwarfs we presently observe around the MW with appropriate subhaloes produced in dark matter only (DMO) simulations based on their properties~\citep[see, e.g.][]{2010MNRAS.408.2364S,2011MNRAS.415L..40B,2013MNRAS.431.1366S}. However this poses two problems. There is an over-abundance of low-mass subhaloes \citep[``The Missing Satellites'' problem;][]{1999ApJ...522...82K,1999ApJ...524L..19M,2010arXiv1009.4505B} and an over-abundance of larger subhaloes with $V_{\rm max} \geqslant 25$km\,s$^{-1}$ ~\citep[the ``Too-Big-To-Fail'' problem;][]{2011MNRAS.415L..40B,2012MNRAS.422.1203B}. With more UFDs being continually discovered thanks to surveys like the Dark Energy Survey~\citep[see e.g.][]{2015ApJ...807...50B,2015ApJ...813..109D,Koposov2015}, the number of Milky Way satellites may soon be broadly consistent with the predictions of DMO simulations. However, resolving the ``Too-Big-To-Fail'' and ``Missing Satellites'' problems requires knowledge of both the dark halo masses of these newly identified objects and of the detailed physics of galaxy formation in low-mass haloes. Such insight can only be obtained through simulations capable of resolving the feedback from individual star formation events. Comparing observable satellites to those produced in simulations is likely a much more complicated issue which depends heavily on other factors such as environment, reionization and the stellar IMF at very high redshifts ~\citep[z\,$\sim$\,15; ][]{arXiv:1406.7097}, all of which are poorly resolved in cosmological simulations and completely neglected in DMO simulations. 

A third issue with comparison to DMO simulation is the fact that the observed rotation curves of many gas rich dwarf (low surface brightness) galaxies show evidence of cored dark matter density profiles \citep[e.g.][]{2006ApJS..165..461K,2008AJ....136.2761O,2001ApJ...552L..23D,2013MNRAS.433.2314H}, rather than the cuspy profiles expected from dark matter only simulations~\citep{1996ApJ...462..563N,1997ApJ...490..493N}. The situation for dSphs is less clear as stars are the only kinematic tracers of the potential in the absence of gas. The literature contains claims of both cores~\citep[e.g.][]{Battaglia08,2011ApJ...742...20W,2011AJ....142...24O,2012MNRAS.419..184A,2012ApJ...754L..39A} and cusps~\citep[e.g.][]{2010MNRAS.408.2364S,2014arXiv1406.6079S,2014MNRAS.441.1584R} as well as suggestions that there is no clear evidence for either from the current available data~\citep[e.g.][]{2013A&A...558A..35B,2014ApJ...791L...3B,2013ApJ...775L..30J}. Several authors have shown that in larger satellites ($M_{\mathrm{vir}}\geqslant 10^9 M_{\odot}$) with bursty SFHs, baryonic effects can transform cusps into large $\sim$1kpc cores~\citep{2002MNRAS.333..299G,2005MNRAS.356..107R,2010Natur.463..203G,2012MNRAS.421.3464P,2012ApJ...761...71Z}. 

Notwithstanding the issue of cusps versus cores, it is well established that feedback from massive stars interacting with the interstellar medium (ISM) has a significant effect on the evolution of dSphs due their low masses and shallow gravitational potential wells~\citep[]{1974MNRAS.169..229L,1986ApJ...303...39D,2010Natur.463..203G,2006MNRAS.371..885R}. Kiloparsec-scale ``superbubbles'' and filaments have been observed in star-forming dwarfs~\citep[e.g.][]{1997ApJ...475...65H,1996ApJ...465..680M,1995ApJ...438..563M,2001ApJ...554.1021H,2005MNRAS.358.1453O} with expansion speeds of $30-60\,$km\,s$^{-1}$. Evidence for these outflows can also be found in the high abundance of metals in the intergalactic medium~\citep[e.g.][]{1995AJ....109.1522C, 1996AJ....112..335S, 2003ApJ...596..768S} and the low abundance of metals contained in dSphs compared to that expected to be synthesised in SN. For example, observational studies by~\cite{2011ApJ...727...78K,kirby11} found that 8 Milky Way dSphs had lost 96 per cent to 99 per cent of the metals produced by their stars.

The role of supernova (SNa) feedback on dwarf galaxy evolution has been investigated by several authors. \citet{2010MNRAS.402.1599S} produce dwarfs similar to those observed in the local group by simulating isolated halos with SN feedback and UV heating only. In order to produce dSph analogues they find that an injection of energy is required to eventually shut off star formation and remove all gas by z=0. An additional external mechanism such as UV radiation can shut off star formation much earlier, reproducing those dSphs with only a single burst of star formation. Similarly, \citet{2009A&A...501..189R} and~\citet{2012A&A...538A..82R} simulate isolated dwarfs and find they still contain a significant amount of gas after $\sim14$ Gyr for halo masses in the range $10^{8}-10^{9}M_{\odot}$. 

In this paper, we highlight an issue which has not yet been satisfactorily discussed in the literature - namely that galaxies like the Ursa Minor dSph, which formed the majority of its stars more than 10 Gyr ago, were actively star forming for an extended period~\citep[$\sim$\,3Gyr;][]{2015MNRAS.449..761U} before their haloes grew to masses above $10^8$M$_\odot$. Thus, the resolution limits of current cosmological simulations mean that they are unable to resolve the physical process in a dSph like Ursa Minor \textit{during the epoch when it formed the bulk of its stellar population}.

The chemistry of the stellar populations requires enrichment of the star forming gas over time: gas retention is thus a key requirement for a successful model of early dSph evolution. Many previous numerical studies have found that haloes below $\sim5\times10^7$M$_\odot$ form few or no stars either due to internal effects~\citep[e.g. SNa feedback: ][]{2016MNRAS.tmp..504R} or external effects~\citep[e.g. reionisation][]{2002ApJ...572L..23S,2000ApJ...539..517B,2009MNRAS.399L.174O,2016MNRAS.456...85S}. However, these simulations lack the resolution to determine whether the failure of the lowest mass haloes to form stars is a numerical artefact or a physical effect. Further, simulations of the long-term evolution of a dSph must necessarily include important details such as halo mergers and gas accretion. 
In our simulations, we focus our attention on the faintest of the classical dSphs around the Milky Way. We choose to model dSph progenitors with properties based on those of the Ursa Minor and Draco dSphs, which both host a predominantly old stellar population, with almost all stars having an age greater than 10Gyr~\citep[]{2002AJ....123.3199C,2015MNRAS.449..761U}. These dwarfs seem to have experienced only a single, extended burst of star formation.

The structure of this paper is as follows: in section~\ref{sec:simdesc} we describe the initial conditions for our simulations and our treatment of SNa feedback. Section~\ref{sec:results} presents the results of our runs and in section~\ref{sec:discussion} we discuss the implications of these results for our understanding of galaxy formation. Section~5 summarises our conclusions. In appendix~\ref{sec:appstab} we present stability tests which confirm the long-term stability of our initial conditions in the absence of SNa feedback. In appendix~\ref{sec:appconvergence}, we present the results of the convergence tests which we performed to confirm that our main conclusions are not sensitive to our choice of numerical resolution. Finally, in appendix~\ref{sec:appfeedback}, we explicitly compare the initial growth phase of our SNa feedback bubbles to the Sedov-Taylor solution to confirm that we are indeed able to capture the main features of individual SNa events.

\pdfoutput=1
\section{Description of simulations}
\label{sec:simdesc}
All our simulations are performed using a modified version of the N-body plus hydrodynamics code GADGET-2~\citep{2005MNRAS.364.1105S}, with thermal stellar feedback (described in section~\ref{sec:simdesc}), radiative cooling and adaptive time steps. We use the SPHS formalism~\citep{2010MNRAS.405.1513R,2012MNRAS.422.3037R} to properly resolve mixing of the multiphase gas, which develops as a result of the energy injection from SNe, and a second order Wendland kernel to allow for the larger neighbour number required with SPHS while avoiding the pairing instability~\citep{Wendland1995,2012MNRAS.425.1068D}. In our simulations, we use 100 neighbours for the SPH calculations. Both the gas smoothing and softening lengths are adaptive, with a minimum value of 0.4\,pc in the densest regions, and the dark matter particle softening length has a constant value of 2pc. In the majority of our simulations, we use $10$M$_{\odot}$ gas particles and $100$M$_{\odot}$ dark matter particles. However, we also explore the effect of raising or lowering the resolution of the simulations relative to these values to confirm that our conclusions are unchanged by going to higher resolution.

\subsection{Initial conditions}
\label{sec:siminit}
\begin{table*} 
\caption{Summary of the simulations presented in this paper. Each case was simulated using both a spherical and flattened gas distribution. The columns represent the name of the run which we refer to in the text (1), number of SN events (2), Energy per SN event in erg (3), scale radius of the gas distribution for the spherical case ($r_{\rm s}$) and the disk case ($r_{\rm d}$) in kpc (4), disk exponential scale height in kpc at $r_{\rm d}$, ($R = 90$ to $R=110$pc) (5), baryon fraction (6), halo concentration (7), and metallicity (8).}
\label{table:runs}
\centering
\begin{tabular} {lcccccccc}
\hline \hline
Run & No. of SN & Energy per SN & Gas $r_{\rm s,d}$(kpc) & Scale height (kpc) & $f_{\rm b}$ & Halo $c$ & Metallicity ([Fe/H])\\
\hline
Fiducial & 100 & $10^{50}$ & 0.10 & 0.013 & 0.16 & 10 & primordial \\ 
200SN & 200 & $10^{50}$ & 0.10 & 0.013 & 0.16 & 10 & primordial\\
500SN & 500 & $10^{50}$ & 0.10 & 0.013 & 0.16 & 10 & primordial\\
$r_{g}$=50pc & 100 & $10^{50}$ & 0.10 & 0.011 & 0.16 & 10 & primordial\\ 
$r_{g}$=200pc & 100 & $10^{50}$ & 0.10 & 0.019 & 0.16 & 10 & primordial\\ 
$f_{b}$=0.08 & 100 & $10^{50}$ & 0.10 & 0.017 & 0.08 & 10 & primordial\\
$f_{b}$=0.04 & 100 & $10^{50}$ & 0.10 & 0.019 & 0.04 & 10 & primordial\\ 
C=4 & 100 & $10^{50}$ & 0.10 & 0.016 & 0.16 & 4 & primordial\\ 
C=30 & 100 & $10^{50}$ & 0.10 & 0.010 & 0.16 & 30 & primordial\\ 
{[Fe/H]=-2} & 100 & $10^{50}$ & 0.10 & 0.013 & 0.16 & 10 & -2 \\
{[Fe/H]=-1.5} & 100 & $10^{50}$ & 0.10 & 0.013 & 0.16 & 10 & -1.5 \\
\hline
\end{tabular}
\end{table*}

Table~\ref{table:runs} summarises the initial conditions for our suite of simulations. Since our goal is to explore the impact of SNa feedback produced by the first burst of star formation in a dSph progenitor, we require that our initial conditions reflect the likely properties of dSph haloes at that epoch. Observations of the Ursa Minor dSph~\citep[see e.g.][]{2015MNRAS.449..761U} show that it formed the bulk of its stellar population over a period of approximately $2-3$Gyr, starting at around z\,$\sim$\,10 when the age of the Universe was $\sim 0.5\,$Gyr. A number of authors have found that the present-day properties of Ursa Minor are consistent with its occupying a $10^9\,$M$_\odot$ halo~\citep[e.g.][]{1998ARA&A..36..435M,2010MNRAS.406.1220W}. In order to estimate the z\,$\sim$\,10 progenitor mass for a halo with a dark matter mass of $10^{9}\,$M$_{\odot}$ halo at $z=0$ we considered the mass assembly history of dark matter haloes.  We used the ~\citet{2008MNRAS.383..557P} algorithm to generate several thousand realisations of Monte Carlo merger trees of low-mass haloes, fixing their virial mass at z=0, and determining the distribution of masses of their main progenitors at z=10 (and 6). We find a mass of $\sim3\times10^{7}\,$M$_{\odot}$ at $z\sim10$, which agrees with estimations from other studies~\citep[e.g.][]{2010MNRAS.406.2267F, 2010MNRAS.402.1599S}.

As we treat our dSph as an isolated system, external effects that will increase its size and mass are not included. The merger history of such a halo therefore sets an upper limit to the duration over which our simulation is meaningful. It is reasonable to assume that this halo mass would remain roughly constant for around 1Gyr (starting at z=10 and ending at z=4), increasing by at most a factor 3-4~\citep[][]{arXiv:1406.7097}. As the halo grows, mass will mostly be accreted at large radii~\citep[]{2003ApJ...597L...9Z} and so the impact on the central regions where the stars and gas are located, will be minimal.

In this study, we implicitly assume that a $3\times10^7$M$_\odot$ halo at $z\sim10$ would have had sufficient time to accrete $\sim5\times10^6$M$_\odot$ of gas and that this gas would have been able to cool to the densities in our initial conditions. The uncertain gas acquisition rates make it difficult to verify this assumption. However, given that gas in this halo has a virial temperature of a few\,$\times10^3$K at z\,$\sim$\,10, the cooling time for low metallicity gas would be $\lesssim 10^6\,$yr~\citep[see eqs. 8.93 and 8.93 of][]{MovdBWhite2010}. For comparison, the age of the Universe at $z\sim 10$ is $\sim0.5\,$Gyr. Thus, in the absence of external heating, there is certainly sufficient time for the dSph to have begun forming stars provided it had already accreted gas. Cosmological simulations are only beginning to approach the resolutions required to study the gas accretion histories of haloes below $10^8\,$M$_\odot$~\citep[see e.g.][]{sawala2015apostle,2015MNRAS.452.1026L}, so a fully consistent treatment of this aspect of the problem is outside the scope of this paper. We note, however, that~\cite{2016MNRAS.tmp..504R} found that even a $10^7$M$_\odot$ halo was able to form stars in their simulations, indicating that gas cooling is unlikely to be a challenge for such haloes.

Our fiducial model is based on the observed properties of the Draco and Ursa Minor dSphs. Both galaxies formed $\sim3\times10^{5}$M$_{\odot}$ of stars over $\lesssim3$Gyr~\citep[][]{2005astro.ph..6430D,2002AJ....123.3199C}. Assuming a cosmic baryon fraction of 0.16, our dSph halo contains an initial gas mass of $4.8\times10^{6}M_{\odot}$. If the entire stellar population of the dSph formed from this gas, this would correspond to a star formation efficiency of $\sim6.5$ percent - clearly this assumes that later accretion events increase the dark matter mass without adding to the gas content of the halo. For a Salpeter stellar initial mass function~\citep[IMF;][]{1955ApJ...121..161S},  the formation of a stellar mass of $3\times10^{5}M_{\odot}$ would generate $\sim2000$ SN events (where we have assumed a mass range of $8-20$M$_\odot$ for the SNa progenitors, giving $\sim2\times10^4$M$_\odot$ in SNa progenitors with a mean mass of $12$M$_\odot$). As our simulations cover a period of 1Gyr we would expect approximately $10^3$ SNe to occur over the course of the simulation. Even if each SNa deposited only 10 per cent of its total energy output (i.e. $10^{50}\,$erg) into the gas, this would exceed the total binding energy of the gas in any of our simulations by at least a factor of two (the maximum binding energy of the gas in our simulations is $\sim5\times10^{52}\,$erg). However, our initial experiments showed that comparing the gas binding energy to the energy input from the SNe significantly under-estimated the gas loss from a low-mass halo as we found that the gas became unbound before the energy injected reached that of the binding energy. We reduce the number of SN events to explore the amount of energy that can be injected into the ISM before the gas becomes too hot and extended to continue to build up the rest of the stellar population over the next $1-2$\,Gyr following on from our simulations. Our fiducial model includes only $100$ SNe, although we also carry out some runs with $200$ or $500$ SNe for those initial configurations which we find to be more resilient. We return to this issue in Section~\ref{sec:discussion}.

\subsubsection{Dark Matter halo}
In all our simulations, the dark matter halo follows a Hernquist profile~\citep{1990ApJ...356..359H}
\begin{equation}
\rho(r)= \frac{M_{200}}{2\pi}\frac{a_{\rm h}}{r(r+a_{\rm h})^3}
\end{equation}
where $a_{\rm h}$ is a scale radius defined by the concentration parameter from an NFW profile~\citep{1996ApJ...462..563N}, assuming the two profiles contain the same mass within $r_{200}$. 
The positions and velocities for the halo particles were generated allowing for the potential of the gas and stellar components using the codes mkgalaxy and mkhalo~\citep{2007MNRAS.378..541M} within the NEMO~\citep{1995ASPC...77..398T} environment. In all simulations $M_{200} = 3\times10^7M_{\odot}$. We take the halo concentration to be $c=10$ in all but two runs, which is a reasonable value for for halos at $z\sim10$ \citep[]{2011ApJ...740..102K} and corresponds to the maximum concentration value at $z=4$ where our simulations end. These values imply an NFW scale length of $0.3\,$kpc and Hernquist parameters of $a_{\rm h}=0.5$\,kpc and $r_{200}=3\,$kpc. 

We re-run the fiducial simulation at higher and lower concentrations to investigate the effect on the resulting dwarf. Although we do not expect a dwarf progenitor to have such a high concentration at $z=10$, we include $c=30$ to compare with other work in which the concentration is kept constant for the lifetime of the dwarf~\citep[]{2016MNRAS.tmp..504R} at present day values. 
Cosmological simulations find that all halos converge to a concentration of 4 at high redshift ~\citep[$z>8$;][]{2003ApJ...597L...9Z,2009ApJ...707..354Z,2011MNRAS.411..584M}, although these studies do not include the mass scales we are interested in due to resolution constraints. However,~\citet{2015MNRAS.452.1217C} use halo mass assembly histories to investigate halo concentrations down to much smaller mass scales and find that at z=10 the concentration of all halos in their simulations are in the range $c\sim3$ to $c\sim4.5$. We therefore consider a value of $c=4$ as an example of a low concentration value our progenitor dSph haloes.  

\subsubsection{Baryonic components}
In order to explore the effect of gas morphology on the ability of the dSph to retain its gas, we consider both
spherical and disc morphologies for the initial gas distribution. In the spherical case the gas and stellar components are distributed in a smooth Hernquist profile (using the method described above for the halo) with a scale radius of 100pc, which makes both components more concentrated than the dark matter halo. The total gas and stellar masses are $4.8\times10^{6}$M$_{\odot}$ and $10^{4}$M$_{\odot}$ respectively, giving particle masses of $10$M$_{\odot}$ for both the gas and stars. The gas follows an ideal equation of state and is assumed to be in hydrostatic equilibrium which determines the temperature profile. Having generated our initial conditions, we evolve them for 300\,Myr excluding stellar feedback, to allow the components to settle fully into equilibrium (see Appendix~\ref{sec:appstab}).

The gas disc was created using the code 'DiscGO' which creates equilibrium disc galaxies \citep{2013MNRAS.434.3606N}. Both gas and stellar components have the same scale length and follow an exponential surface density profile:

\begin{equation}
\Sigma_{g}(R)=\frac{M_{g}}{2\pi {r_d^2}}e^{-R/r_d}
\end{equation}

The gas is assumed to be in hydrostatic equilibrium which determines the height of the gas disc. The total gas mass, stellar mass and the number of particles of each are equal to that used in the spherical case. As with the spherical case we run the initial conditions with no feedback to allow the components to settle into equilibrium. Figure~\ref{fig:diskstab} shows the disc initial conditions running for 1.25Gyr to confirm that they are indeed stable. The distribution at 300\,Myr is used as the initial conditions for the run with feedback, as our stability tests showed that the initial settling of the components due to numerical fluctuations and the IC generation process had been completed by this time.

\subsection{Cooling and Feedback}
Modelling the evolution of a supernova remnant (SNR) is computationally demanding due to the large dynamic range of spatial scales over which the physics needs to be resolved. Both cosmological simulations and simulations of single galaxies lack the resolution to accurately model individual SN events and use approximations to inject energy into the ISM from a single star particle representing a stellar population. Injecting energy thermally at low resolution results in almost all of the energy being immediately radiated away before it has any effect on the ISM~\citep{2002MNRAS.333..649S,2006MNRAS.373.1074S,2011MNRAS.415.3706C}. This is due to the energy being injected into a large amount of mass (giving the particles lower temperatures, and so shorter cooling times) and the inability of the particles to react quickly to this change due to the time resolution. 
Solutions to this problem vary. They include delaying radiative cooling by hand \citep[e.g][]{2000ApJ...545..728T,2002MNRAS.330..113K, 2003ApJ...596...47S,2004MNRAS.349...52B} or forcing the temperature of the injected particles to be higher to account for the extra energy losses (to make the cooling time longer than the dynamical time) to allow the SNR to expand and interact with the ISM \citep{2012MNRAS.426..140D}. 

A gas particle mass of $10$M$_{\odot}$ enables us to resolve single SN events by injecting energy thermally, giving an isotropic effect on the closest neighbouring gas particles. Thermal energy injection at this resolution gives a good approximation to the Sedov taylor blast wave solution (see appendix~\ref{sec:appfeedback}), enabling us to resolve SN without further numerical adjustments. Injecting the SN energy purely as thermal gives a similar approximation to using kinetic energy both on small scales \citep[]{2015MNRAS.451.2757W} and on large scales \citep[]{2012MNRAS.426..140D,2012MNRAS.419..465D} provided the time stepping used allows the injected particles and their neighbours to respond promptly to the sudden energy input. Each SN event injects $10^{50}$ erg of thermal energy into $\sim100$ neighbouring gas particles. The amount of energy each particle receives is kernel-weighted by its distance from the star particle.



We do not model star formation explicitly, as we are primarily concerned with the ability of our model dSph to retain sufficient gas to support further star formation; therefore the only stars in the simulations are those present in the initial conditions. As discussed in section~\ref{sec:siminit} we would expect $\sim10^3$ SN over the period of time covered by the simulations assuming a single, continuous burst of star formation over $\sim3$Gyr. We include $10^3$ star particles, a number of which are chosen to inject energy at random times during the course of the simulation, depending on the run. For example, in our fiducial run we have 100 SNe which translates to a rate of 1 SN every 10\,Myr. 

Radiative cooling of the gas is included following the method of~\cite{1996ApJS..105...19K} down to $10^4$K, assuming ionisational equilibrium. We further assume that the gas is of primordial composition and do not model metals produced by supernovae in these simulations. As gas cooling is dependant on metallicity we investigate the effect the presence of metals has on our results by repeating the fiducial run for both the spherical and disk morphologies at two different metallicities: [Fe/H]$=-1.5$ and $-2$.  To account for cooling due to metals we use approximate cooling functions based on those described in~\cite{1993ApJS...88..253S} for temperatures $10^{4}-10^{8.5}$K. Cooling below $10^{4}$K is modelled as described in~\citet{2008Sci...319..174M} via the fine structure and metastable lines of heavy elements. These cooling curves were applied to all of the gas in the simulation. This is therefore an extreme case, as the majority of metals would be those ejected by the SNe which would not be present in the initial gas and would be unlikely to be uniformly distributed through the whole dSph. The results are presented in Section~\ref{sec:metallicity}.

We neglect external processes in the simulations such as mergers and the effect of heating from the cosmic UV background (UVB). Observations of the IGM indicate that reionization of the universe was complete by $z\sim6$~\citep[see e.g.][]{2001AJ....122.2850B,2003AJ....125.1649F}. It has been suggested by several authors that reionization could alleviate the missing satellite problem by suppressing star formation in low mass haloes or preventing them from forming stars altogether~\citep[see, e.g.][]{2000ApJ...539..517B,2015MNRAS.448.2941S}. However dSphs like Ursa Minor and Draco were continuously forming stars for $\sim3$Gyr (up to $z\sim3$) suggesting that reionization was not responsible for ending star formation in these dwarfs. It has been shown that the effects of self-shielding at high redshift are quite significant, and baryons in high density peaks are still able to cool and form stars~\citep{2004ApJ...600....1S,2005ApJ...629..259R}. Therefore those low-mass haloes that are able to maintain a high density of gas in their centres during an episode of star formation could possibly continue to form stars even after the universe has been reionized.

Our simulations run for 1Gyr so the effects of the UVB would only be relevant at the end of our simulations and would only act to maintain the temperature of the already hot and extended gas that has been ejected from the galaxy due to the SNe feedback.

\section{Results}
\label{sec:results}
As noted in the Introduction, observations of the Milky Way satellites provide evidence that they experienced extended periods of star formation which in turn indicate that they were able to retain their gas, and continue star formation, until Type~Ia SNe had had time to enrich the gas with iron. Thus, a minimal requirement for a successful model of a dSph progenitor is that it be able to retain gas in its inner regions at sufficiently high densities and low temperatures to support further star formation.

We begin by exploring the impact of initial gas morphology on the ability of a dwarf galaxy progenitor to retain its gas in the presence of feedback from SNe. Sections~\ref{sec:spherical} and~\ref{sec:disk} present the results from the spherical and disk gas morphologies separately: in each case we discuss the impact of each of our model parameters on the resulting evolution, under the assumption that all the gas has a metallicity of [Fe/H]=-3. In Section~\ref{sec:metallicity} we explore the impact of a higher initial gas metallicity on our conclusions. 

\subsection{Spherical gas distributions}
\label{sec:spherical}
\begin{figure}
\includegraphics[width=0.5\textwidth]{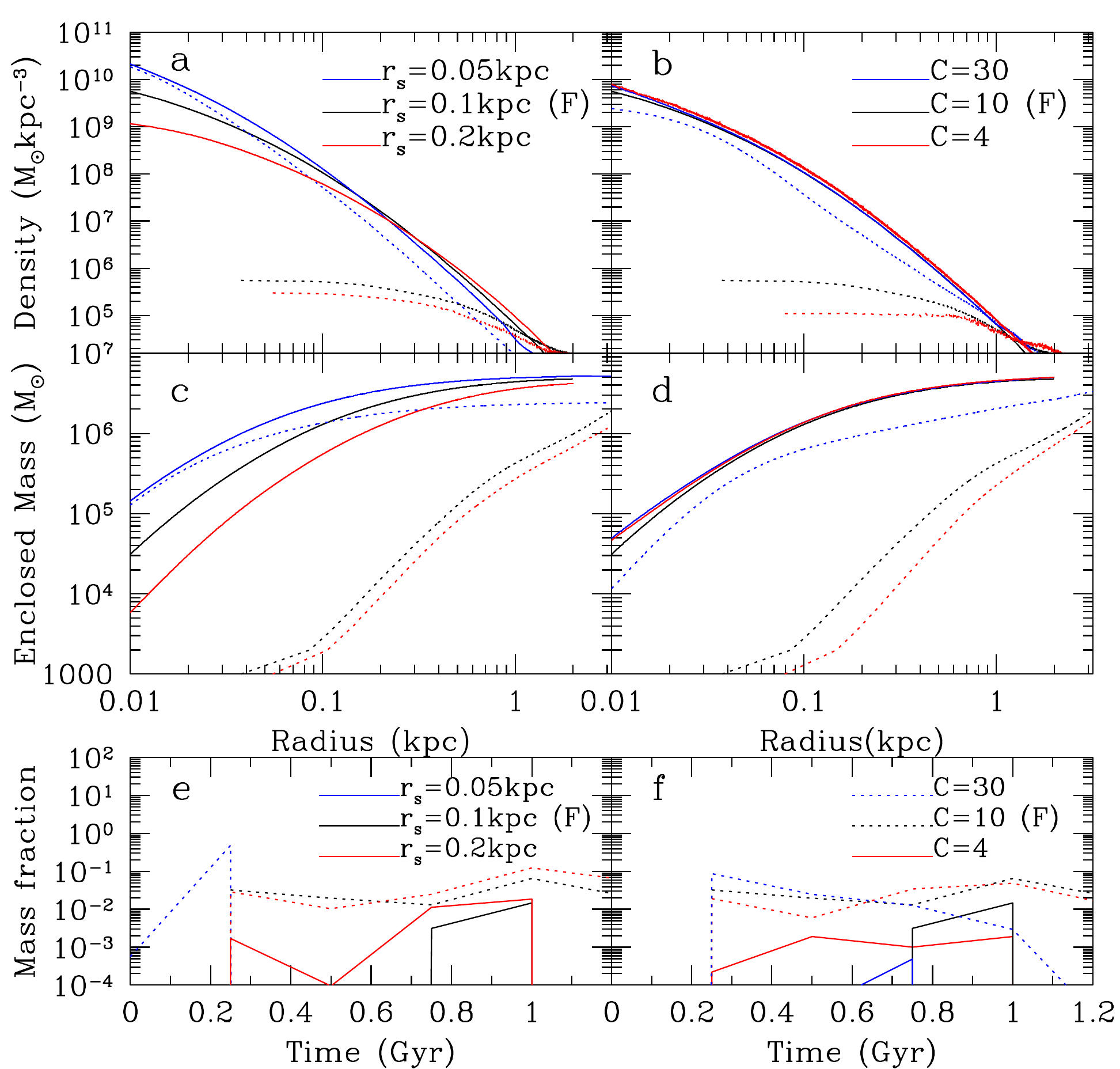}
\caption{\textbf{Top, Middle:} The evolution of the gas density (top) and enclosed mass (middle) profiles for simulated dSphs with initially spherical gas distributions. The solid curves show the initial profiles and the dashed curves show the resulting profiles after 1.25\,Gyr. The black curves correspond to our fiducial initial conditions, while the red and blue curves indicate changes to either the scale length of the gas distribution (left panels) or the concentration of the dark matter halo (right panels). The gas particles are binned radially with 100 particles in each bin. \textbf{Bottom:} The fraction of gas with velocities greater than the local escape velocity as a function of time for the simulations in the upper panels. Solid (dashed) curves give the fraction of gas at radii less than 1\,kpc (1 $\leqslant r \leqslant$ 5\,kpc). (F) denotes the fiducial run.}
\label{fig:sphere_gas} 
\end{figure}
\begin{figure}
\includegraphics[width=0.5\textwidth]{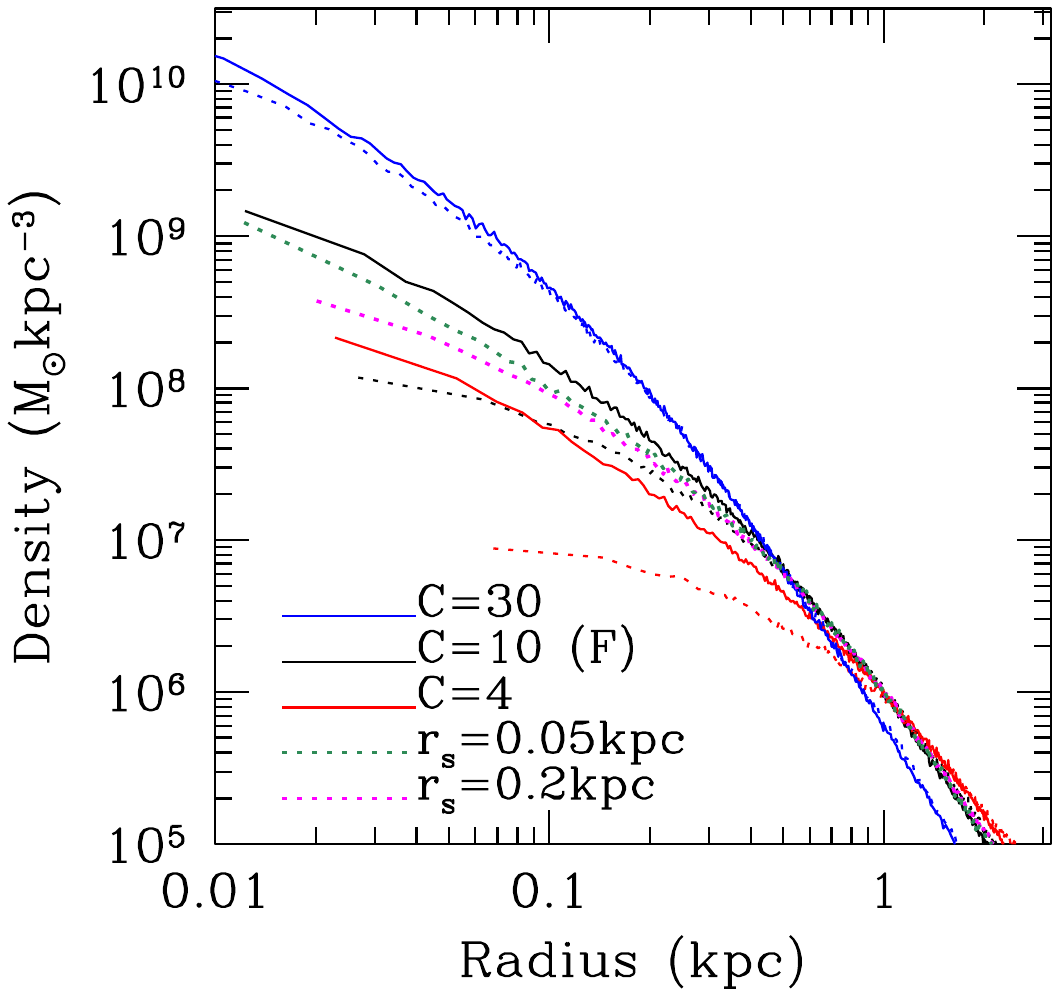}
\caption{Evolution of the dark matter halo density profiles for the simulations presented in Figure~\ref{fig:sphere_gas}. For clarity, we show the initial density profiles for the fiducial ($c=10$, solid black curve) , $c=4$ (solid red curve) and $c=30$ (solid blue curve) simulations only, as the initial profiles for both other models shown are indistinguishable from the fiducial model. The profiles after $1.25$\,Gyr of evolution are shown as dashed curves with the colours indicating the model. The halo particles are binned radially with 300 particles in each bin. (F) denotes the fiducial run.}
\label{fig:sphere_halo} 
\end{figure}

The black curves in Figure~\ref{fig:sphere_gas} show the evolution of the gas density (top panels), the enclosed gas mass (middle panels) and the fraction of gas at the escape velocity (bottom panels) for simulations in which the gas distribution is initially spherical and assuming our fiducial (F) parameters. In the top four panels, the solid curves show the initial distribution and the dashed curves correspond to distributions after 1.25\,Gyr. Figure~\ref{fig:sphere_gas}(a) shows that by $1.25$\, Gyr the gas density has been significantly reduced throughout the dSph, with the central gas density being more than four orders of magnitude lower than its initial value. The dashed black curves in Figure~\ref{fig:sphere_gas}(c) confirm that that by $1.25$\,Gyr more than 90 per cent of the gas has been moved beyond a radius of 1kpc. We have not included the results from the simulations with 200 or 500 SNe (the fiducial run has 100SNe) as the density and enclosed mass are simply reduced even further in these models. 

Clearly, the energy injected by 100 SNe is sufficient to re-arrange the bulk of the gas in this simulation. This is not surprising as the total binding energy of the gas in the central 100\,pc is only about 20 per cent higher than the $10^{52}$\,ergs of thermal energy injected by 100 SNe in our models. While Figure~\ref{fig:sphere_gas}(e) shows that less than five per cent of the gas at large radii is moving above the local escape velocity, we note that the low density of this gas (more than two orders of magnitude below the original mean gas density within $1$\,kpc) means both that its cooling time is very long (temperatures are below $10^4$K) and it is very susceptible to removal by any external perturbation, for example ram pressure stripping by the gaseous halo of its host galaxy (at a radius of 100kpc the MW halo would strip any gas outside 10kpc, and at a radius of 10kpc would strip gas outside 2kpc~\citep[]{2013ApJ...770..118M}). It is therefore very unlikely that this gas could be retained by the dSph for sufficient time to allow it to fall back into the inner regions and form additional stars.

The red and blue curves in Figure~\ref{fig:sphere_gas}(a) show the evolution of the gas density profile in simulations with different initial gas concentrations. As expected, it is much more difficult to remove gas from the inner regions of the dSph in the case of a  more concentrated initial distribution ($r_s$ = 0.05\,kpc; blue curves) but somewhat easier to remove gas in the less concentrated case ($r_s$ = 0.2\,kpc; red curves). The corresponding enclosed mass profiles (Figure~\ref{fig:sphere_gas}(c)) show that in the most concentrated run, mass is only removed from the outer parts of the dwarf where the density is lower and cooling times are longer. In contrast to the models with $r_{\rm s}= 0.1$\,kpc or $r_{\rm s}=0.2$\,kpc, which retain less than 0.1 per cent of their gas content within 100\,pc, in the model with $r_{\rm s}=0.05$\,kpc, 15 per cent of the gas initially within 100\,pc remains there after 1.25\,Gyr of evolution. The large differences between the final density profiles arises due to the dependance of cooling time on the gas density: $t_{\mathrm{cool}}\propto\rho^{-1}$. The cooling times in the fiducial case ($r_s$=0.1\,kpc) are already long enough for the gas to be driven to large radii, so further decreasing the density ($r_s$=0.2\,kpc) gives a similar result, whereas for the most concentrated distribution ($r_s$=0.05\,kpc) the cooling times have decreased by a factor $\sim$5 allowing more gas to be retained. 

Overall, Figure~\ref{fig:sphere_gas}(c) shows that a dSph with a more centrally concentrated initial gas distribution loses a larger fraction of its total gas mass: roughly 50 per cent of the gas mass is located beyond 5\,kpc at the end of the simulation. However, the key point is that this dSph retains gas in the inner regions at densities where it has the potential to contribute to further star formation.

The red and blue curves in Figure~\ref{fig:sphere_gas}(b) and (d) show the
impact of dark matter halo concentration on the evolution of the gas density
profile and the enclosed mass profile. As in the case of varying gas
concentration, it is slightly easier to remove gas from a halo with a lower
concentration and significantly more difficult for a halo with a higher
concentration. However, in contrast to the results for the simulation with $r_{\rm
  s}$=0.05\,kpc for the gas initially (the blue curves in
Figure~\ref{fig:sphere_gas}(c)), in the simulation with $c=30$ gas
is depleted all the way to the centre of the dSph, although the fraction of
mass removed from within 100\,pc is similar in both cases (57 per cent and 46 per cent of original mass retained for $r_s=0.05$kpc and $c=30$ respectively). The potential energy of the gas within
30pc in the $r_{\rm s}=0.05$\,kpc and $c=30$ simulations is $1\times10^{52}$ erg
and $5.8\times10^{51}$ erg, respectively. Thus, the significantly different gas fractions within 30pc at $1.25$\,Gyr seen in the two simulations explicitly demonstrates that the ability of a dSph to retain gas in its inner regions is a function not only of the depth of its gravitational potential well but also of the gas density in the inner parts which determines the timescales on which the gas can cool. 

The fraction of gas that reaches the escape velocity is shown in panels (e) and (f) of Figure~\ref{fig:sphere_gas} as a function of time for the same runs. The solid curves correspond to the fraction of gas particles within $1$\,kpc at the time of the snapshot that have velocities above their local escape velocity; the dashed curves show the same fraction for gas particles with radii in the range $1-5$\,kpc. The fraction of high velocity gas within $1$\,kpc is very low in all runs, only briefly reaching 1 per cent in the standard and low (gas or halo) concentration runs. This suggests that the gas remaining within $1$\,kpc after $1.25$\,Gyr will be retained by the dSph in all cases we have considered. However, only in the high (gas or halo) concentration models is the gas density sufficiently high at the end of the simulation that further star formation is likely to occur~\citep[assuming a density threshold of $7.4\times10^{9} M_{\odot}$kpc$^{-3}$ to ensure the gas is dense enough to be mainly molecular;][]{2016MNRAS.tmp..504R}. The gas in the centre of these two dwarfs at the end of the simulation is cool (less than $10^4$K) and dense, and so we would expect star formation to continue.

It is interesting to note that even at radii of $1-5$\,kpc the fraction gas particles with velocities above their local escape velocity is also low, remaining below 10 per cent for most of the simulations. In the $r_s$=0.05\,kpc run, however, it exceeds 10 per cent from about $t=100$\,Myr to $t=400$\,Myr. This is because the initial gas distribution in this simulation has much lower density beyond $1$\,kpc than in the other runs and the outflows from SNe in the inner regions are therefore able to accelerate a large fraction of this gas to velocities above the escape velocity.

Figure~\ref{fig:sphere_halo} shows the response of the dSph's dark matter halo to the evolution of the gas distribution. Following the rapid removal of gas from the centre of the dwarf, the halo central density in the simulation with our fiducial parameter choices (black curves) decreases by an order of magnitude, forming a profile with a much shallower inner log slope within a radius of $\sim$100\,pc. In the simulation initially having less gas within $100$\,pc, due to a less concentrated gas distribution (the $r_{\rm s} = 0.2$\,kpc model; purple dashed curve), the halo profile evolves less dramatically than in the standard model, although the initial halo profiles are indistinguishable. As expected, the removal of the gas from the inner regions in these cases corresponds to a smaller change in the total gravitational potential and therefore the response of the dark matter halo is reduced. The smallest effect is seen when the gas is initially more centrally concentrated (the $r_{\rm s} = 0.05$\,kpc model; green dashed curve) as the retention of a significant fraction of the gas within $30$\,pc offsets the greater gas contribution to the mass budget in this region (gas contributes 44 per cent of the total initial potential within 30pc, compared to 20 per cent for $c=30$). The expulsion of gas to radii beyond $5$\,kpc noted above does not have a significant impact on the dark matter halo.

Only the model with a lower initial halo concentration ($c=4$; red curves in Figure~\ref{fig:sphere_halo}) exhibits more significant halo evolution than the standard model. This is simply because the gas initially constitutes $47$ ($28$) per cent of the gravitational potential within $60$\,pc ($1$\,kpc). The removal of this gas results in the central 100-200\,pc of the dark matter halo evolving to almost uniform density. We will discuss the possible implications of the dependence of the final dark matter halo profile on initial halo concentration in Section~\ref{sec:discussion}.
The halo density profile for the model with the highest halo concentration, $c=30$ experiences no evolution over the run. The potential is dominated by the halo (the gas only contributes 20 per cent in the inner 30pc) and only a small amount of gas is removed, leaving the structure of the halo unaffected.

The simulations presented in this section demonstrate that galaxy progenitors with similar masses to those expected for $z\sim10$ dSphs and with initially spherical gas distributions are generally unable to retain their gas in the face of feedback from just $100$SNe. Progenitors with high concentration haloes are able to retain a higher fraction of gas, but only haloes with highly concentrated gas distributions are able to retain gas at densities which are high enough to sustain star formation.

\subsection{Flattened gas distributions}
\label{sec:disk}
\begin{figure}
\includegraphics[width=0.5\textwidth]{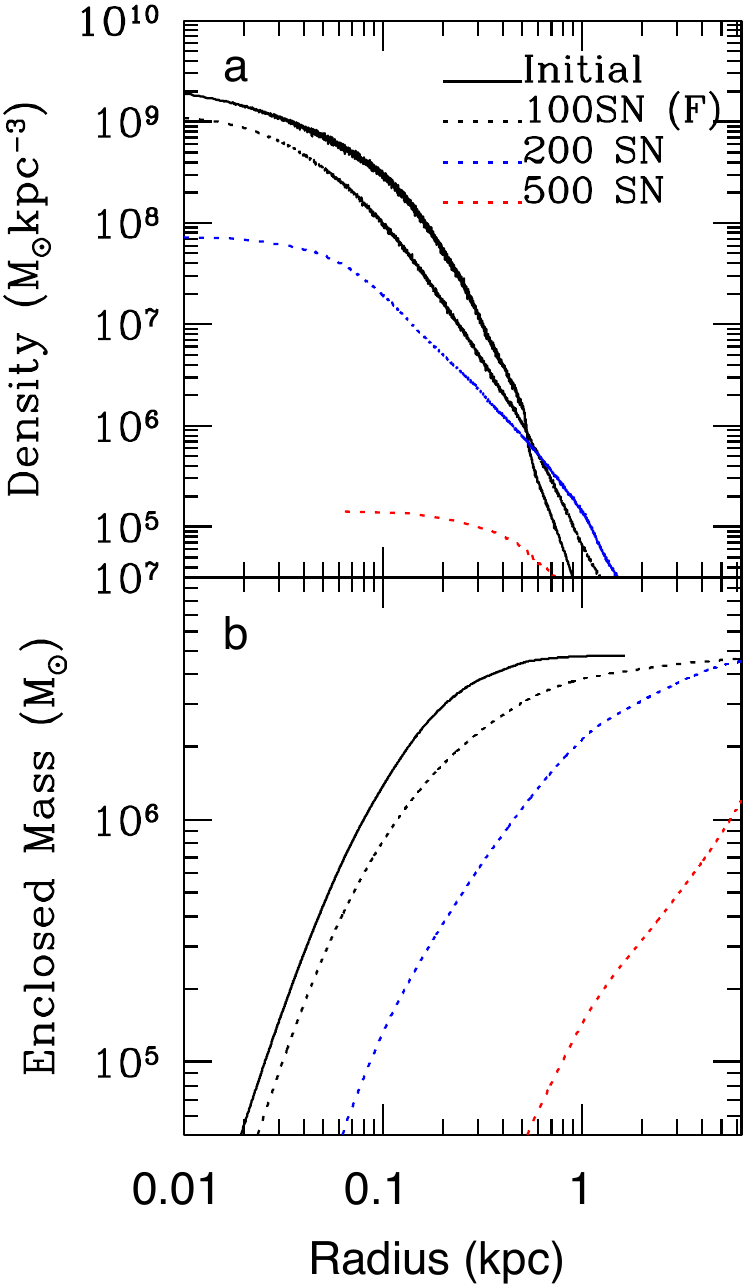}
\caption{Evolution of the gas density (\textbf{top}) and enclosed mass (\textbf{bottom}) profiles for simulated dSphs with gas initially distributed in a disk. In both panels, the solid black curves show the initial profiles, and the black, blue and red dashed lines show the resulting profiles after $1.25$\,Gyr of evolution for simulations in which feedback from 100, 200 or 500 SNe is included. The corresponding SN rates are $0.1$Myr$^{-1}$, $0.2$Myr$^{-1}$ and $0.5$Myr$^{-1}$, respectively. The gas particles are binned radially with 100 particles in each bin and the three dimensional SPH particle density is averaged. (F) denotes the fiducial run.}
\label{fig:disk_gas_1} 
\end{figure}

\begin{figure*}
\includegraphics[width=\textwidth]{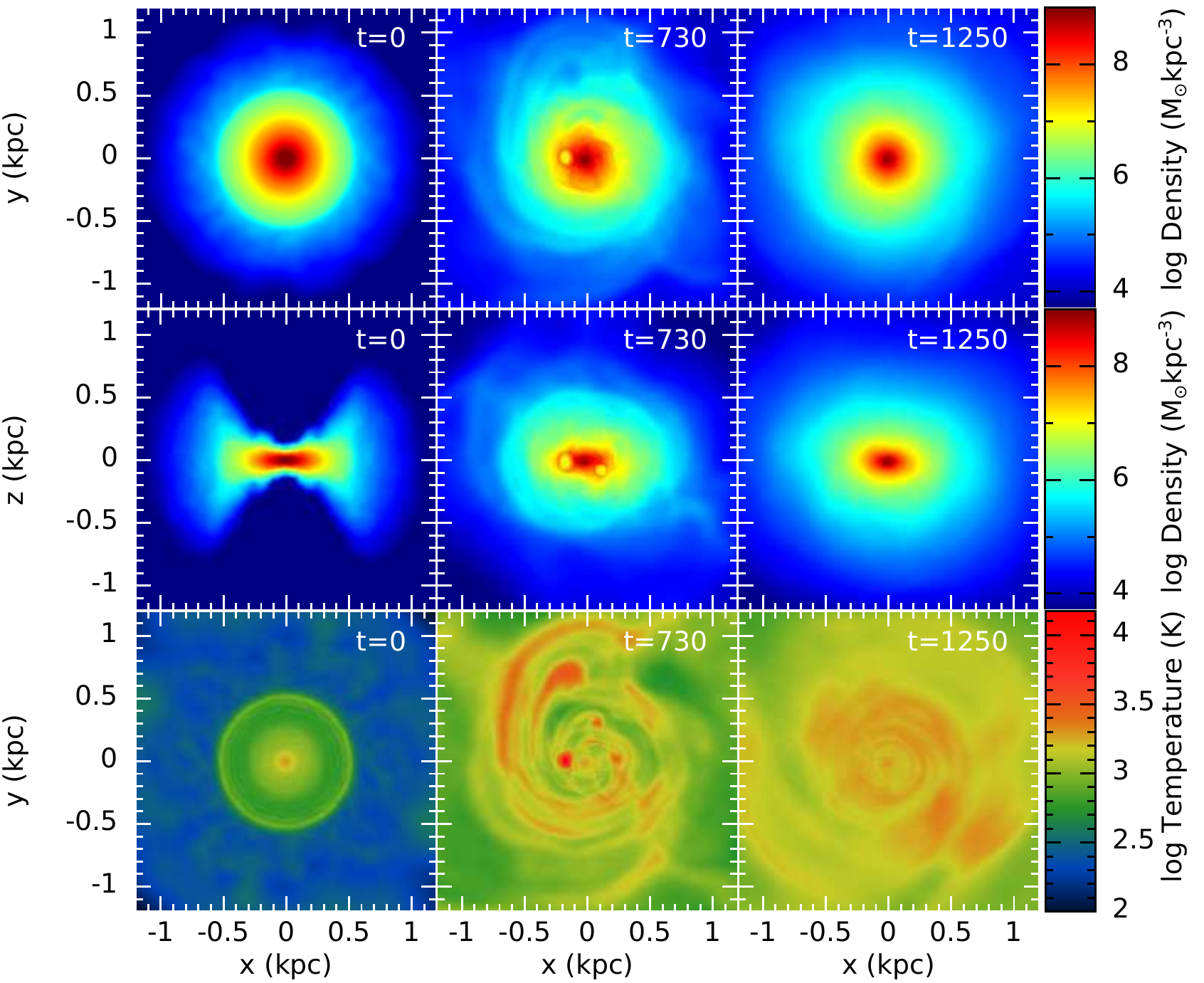}
\caption{Maps showing the evolution of gas density (\textbf{top}, \textbf{middle}) and temperature (\textbf{bottom}) for the fiducial disk simulation. The leftmost column shows the initial conditions, the middle column shows the situation at $730$\,Myr and the rightmost column shows the final state at $1.25$\,Gyr. The quantities plotted are cross-sections showing density in the $x-y$ plane (top panel), density in the $x-z$ plane and temperature in the $x-y$ plane.}
\label{fig:disksplash} 
\end{figure*}

\begin{figure}
\includegraphics[width=0.5\textwidth]{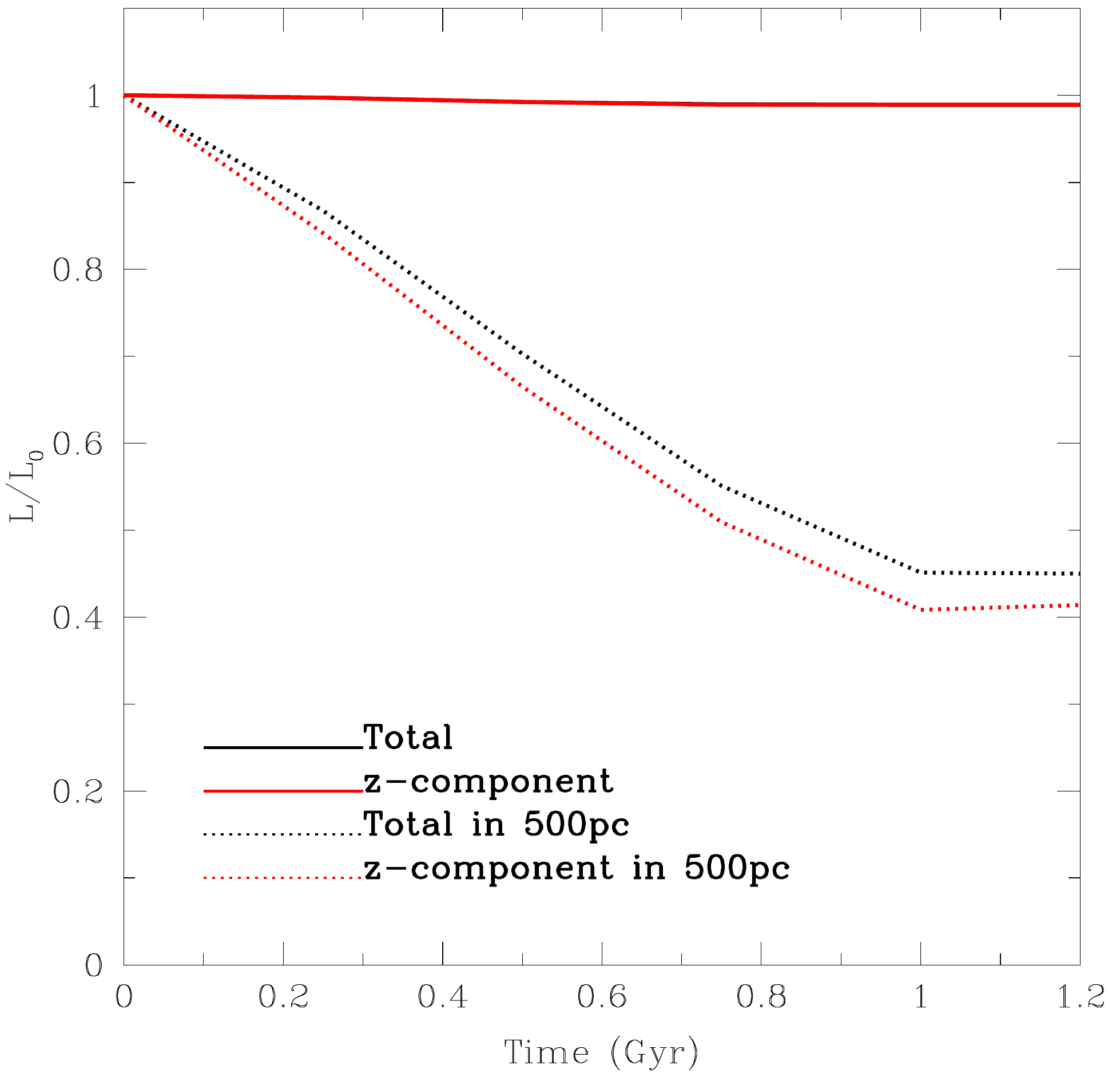}
\caption{The evolution of the total angular momentum (black solid line), and z-component of angular momentum (red solid line) over time for the fiducial disk run compared to the initial values ($L_0$). The same values within 500pc are shown as dashed curves.}
\label{fig:diskangmom} 
\end{figure}

\begin{figure}
\includegraphics[width=0.5\textwidth]{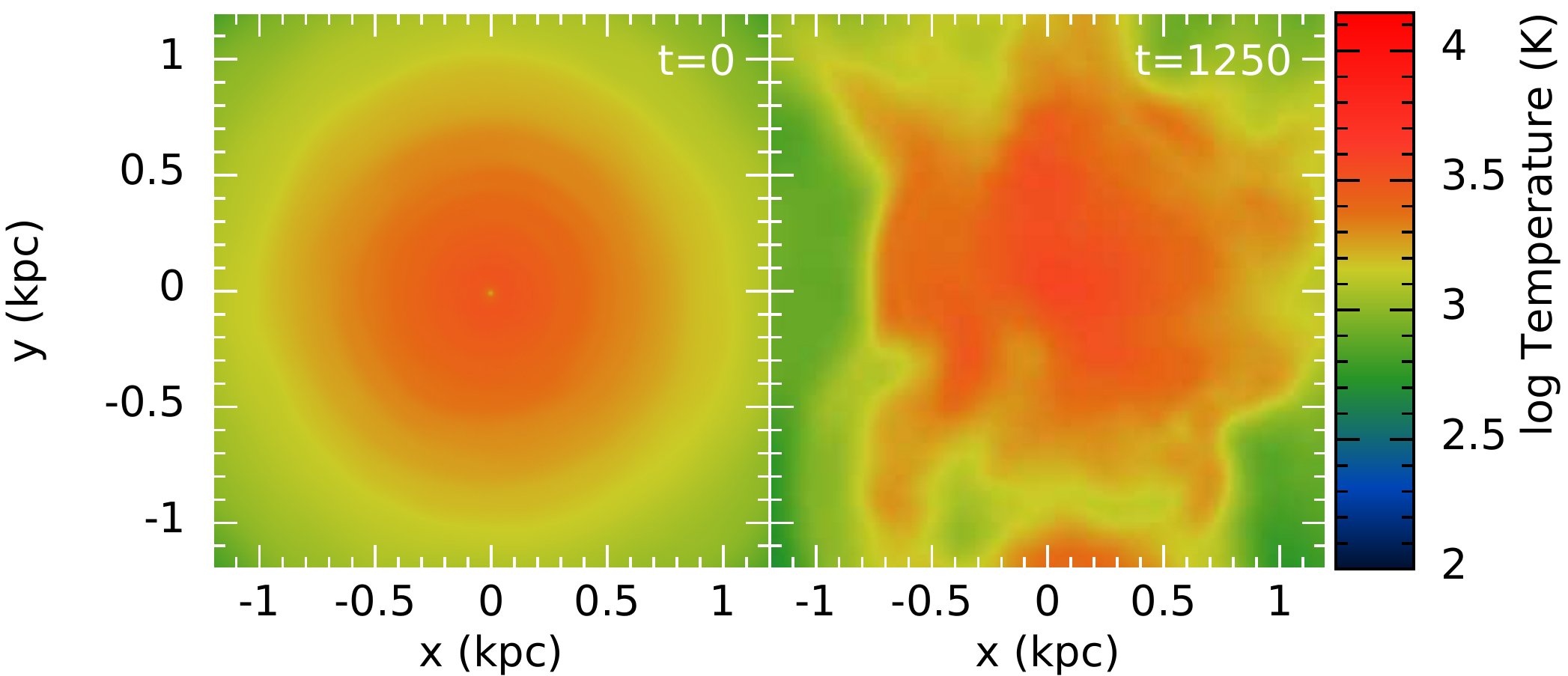}
\caption{Maps showing the initial (\textbf{left}) and final (\textbf{right}:1.25Gyr) gas temperature for the fiducial spherical run in the $x-y$ plane.}
\label{fig:sphere_fiducial_T} 
\end{figure}

\begin{figure*}
\includegraphics[width=\textwidth]{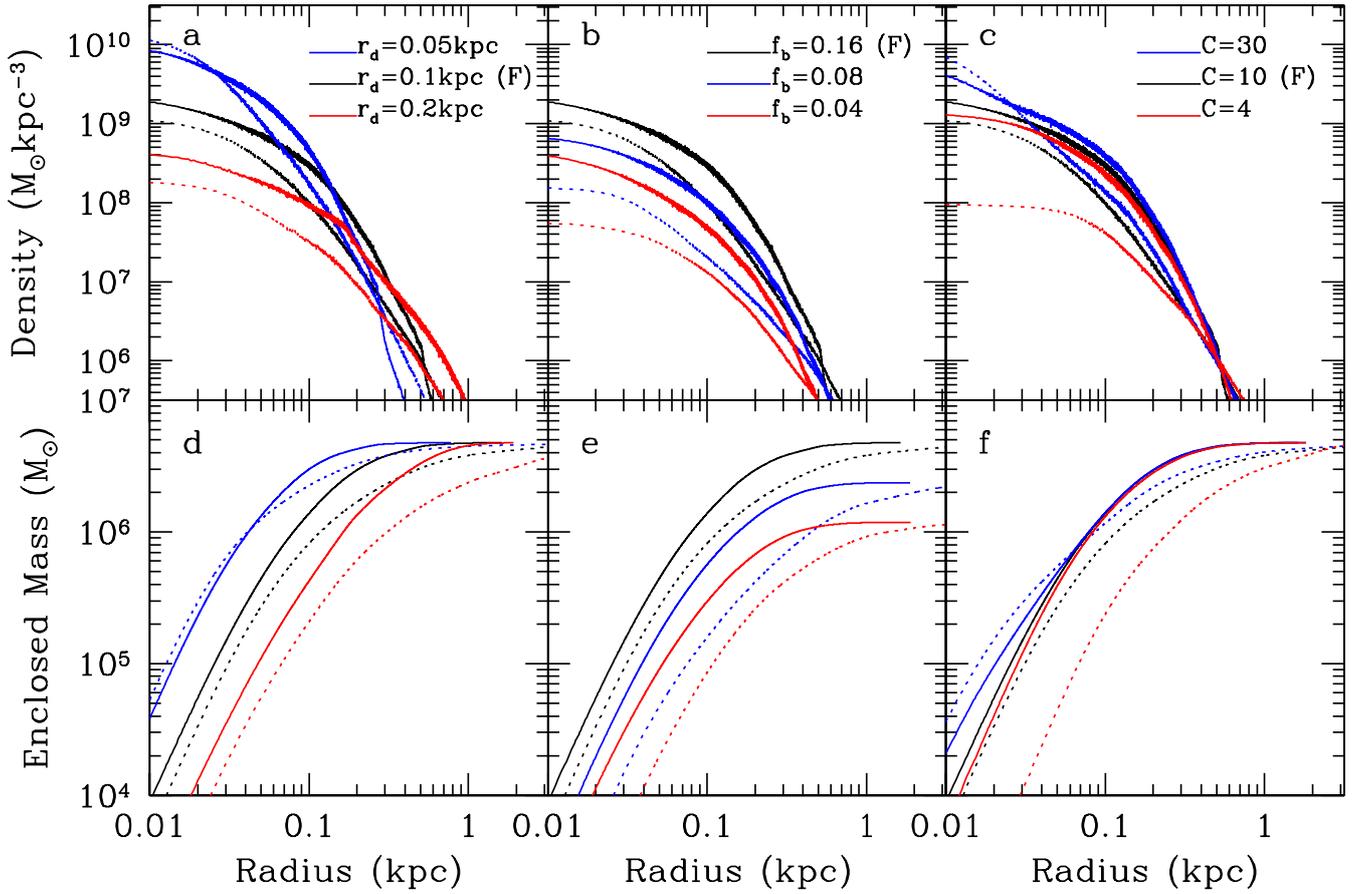}
\caption{Impact on the evolution of the gas density (\textbf{top}) and enclosed mass (\textbf{bottom}) profiles of varying the model parameters for simulated dSphs with gas initially distributed in a disk. In all panels, solid curves denote initial distributions and dashed curves give the profiles after $1.25$\,Gyr of evolution. For ease of comparison, the black curves in each panel show the profiles for the fiducial model, while the red and blue curves correspond to models with alternative values for the initial model parameters. Panels (a) and (d) show the effect of varying the scale length $r_{\rm d}$ of the gas disk; panels (b) and (e) show the effect of varying the total baryon fraction of the initial model; in panels (c) and (f) the dark matter halo concentration is varied. (F) denotes the fiducial run.}
\label{fig:disk_gas_2} 
\end{figure*}

\begin{figure}
\includegraphics[width=0.5\textwidth]{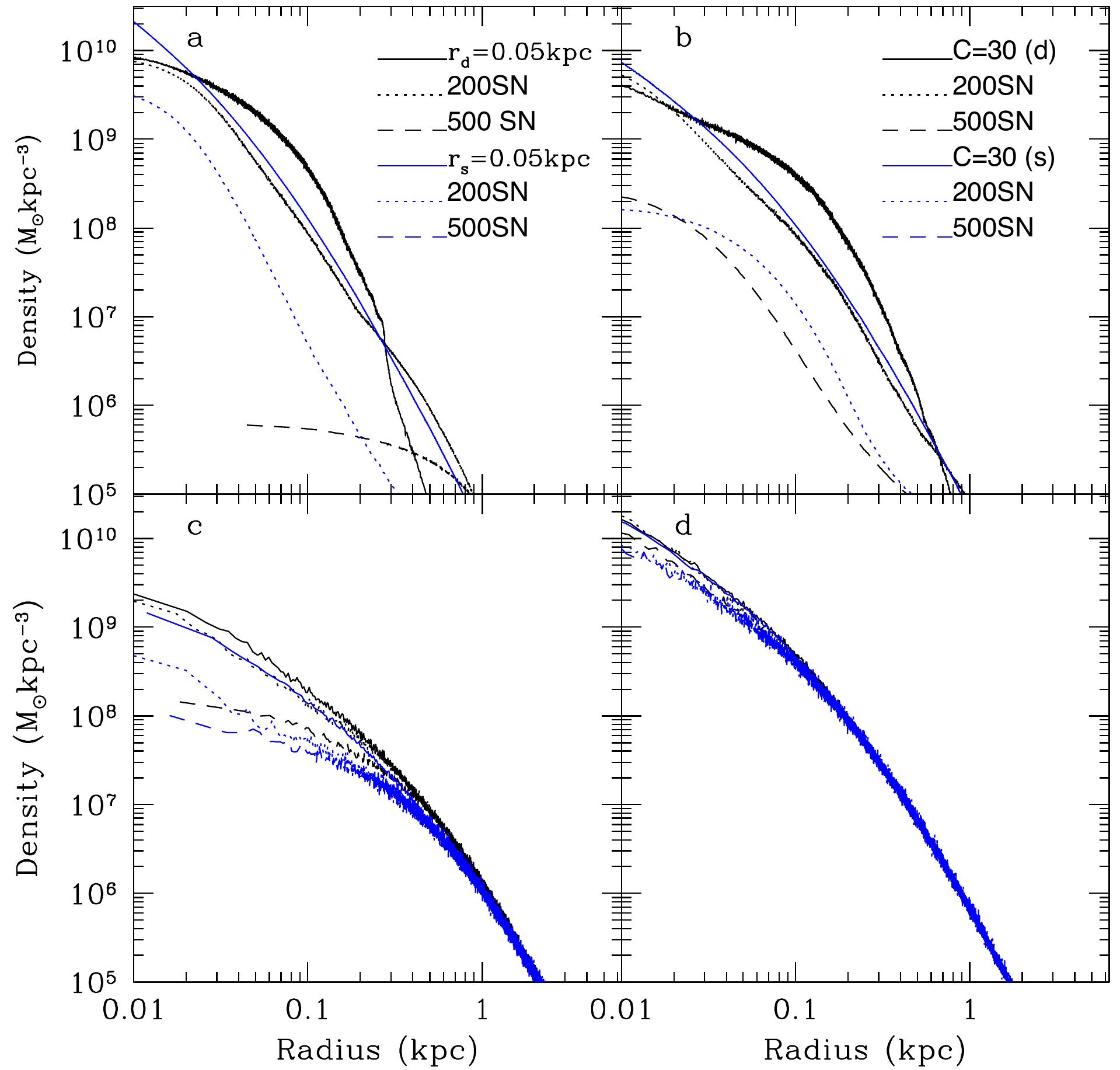}
\caption{Evolution of the gas density (\textbf{top}) and halo density (\textbf{bottom}) profiles for simulated dSphs with a concentrated gas profile (\textbf{left}) and a concentrated halo profile (\textbf{right}). In all panels, the solid black and blue curves show the initial profiles for the disk and sphere respectively, and the dotted and dashed lines show the resulting profiles after $1.25$\,Gyr of evolution for simulations including feedback from 200 and 500 SNe. Note that the gas density curves for the spherical models with 
$500$ SNe (blue dashed curves) fall below the lower limit of the density axis and are therefore omitted for clarity.}
\label{fig:disk_concentration} 
\end{figure}

\begin{figure}
\includegraphics[width=0.5\textwidth]{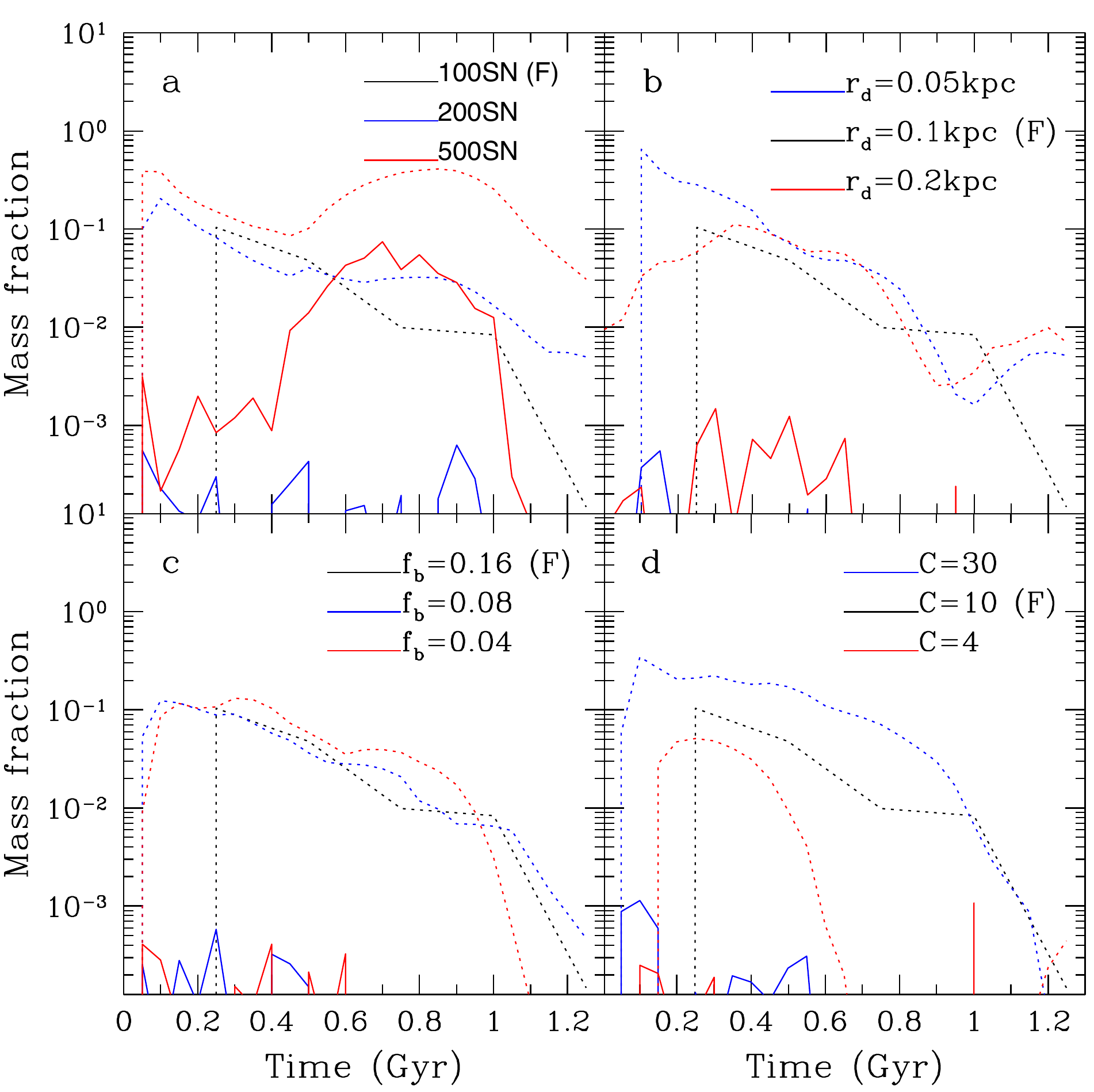}
\caption{The fraction of gas particles with velocities greater than the local escape velocity as a function of time for the simulations in Figures~\ref{fig:disk_gas_1} and~\ref{fig:disk_gas_2}. Solid curves give the fraction of gas particles above escape velocity at radii $r<1\,$kpc while the dashed curves correspond to gas particles at radii $1\,$kpc\,$<r<5\,$kpc.  (F) denotes the fiducial run.}
\label{fig:vesc_disk} 
\end{figure}

\begin{figure*}
\includegraphics[width=\textwidth]{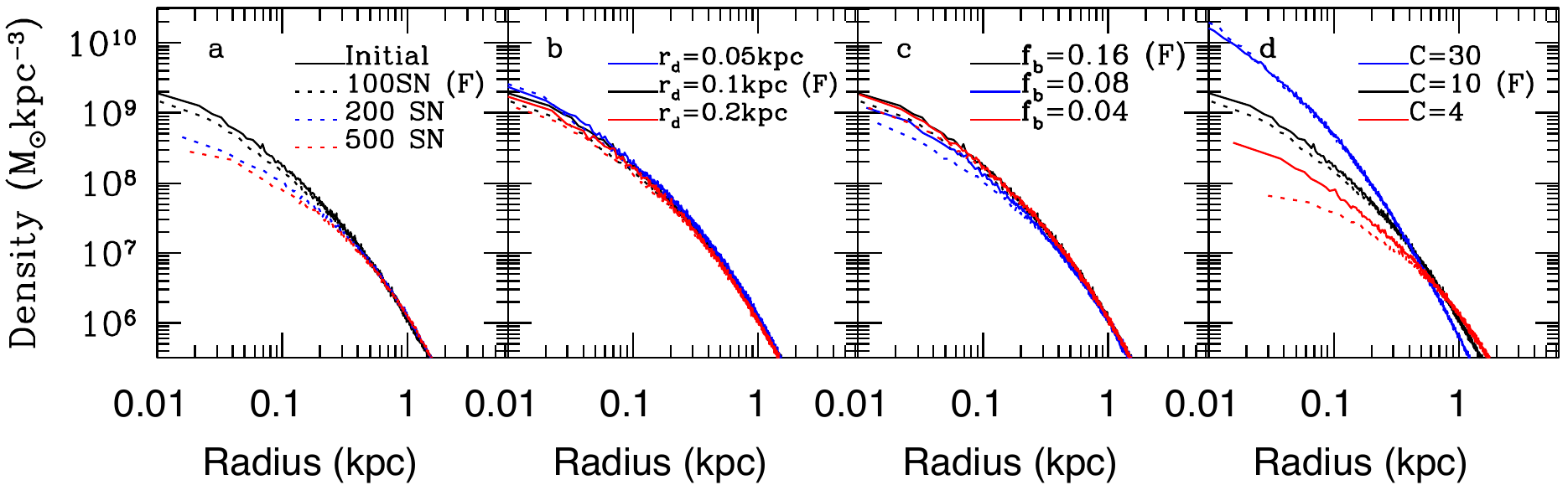}
\caption{Evolution of the dark matter halo density profiles for the simulations presented in Figures~\ref{fig:disk_gas_1} and~\ref{fig:disk_gas_2}. In all panels, the black curves show the profiles for the fiducial disk model. The red and blue curves in each panel show the effect of varying the model parameters: (a) number of SNe; (b) disk scale length; (c) baryon fraction; (d) halo concentration. Solid curves denote initial profiles, while dashed curves show the outcome of $1.25$\,Gyr of evolution. (F) denotes the fiducial run.}
\label{fig:disk_halo}
\end{figure*}

In this section, we consider the evolution of dSph progenitors in which the initial gas distribution is a thick disk ($R_{\rm d}/z_{\rm d}$=1.5 for the fiducial run). Figure~\ref{fig:disk_gas_1} shows that after 100 SN events the disk is able to retain the majority of its gas and the density at the centre remains high enough for further star formation, in the sense that is roughly equal to the highest density thresholds for star formation assumed in simulations~\citep[e.g.][]{2016MNRAS.tmp..504R,2012MNRAS.421.3464P}. This is in contrast to the evolution of initially spherical gas distributions in which 100 SNe are able to push more than 90 per cent of the gas out to radii $>100$pc, whereas the disk retains 80 per cent of the gas initially within this radius. 

The effect of increasing the number of SNe for the disk is also shown in Figure~\ref{fig:disk_gas_1}. The initial conditions are identical for all three runs and are represented by the solid black curve. The dashed black, blue and red curves show the radial density profile after 1.25Gyr for 100, 200 and 500 SN events, respectively. 200 SN events cause the gas density to drop by another order of magnitude, although 60 per cent of the gas still remains within the original 1kpc. As the number of SNe increases, the amount of energy injected exceeds the binding energy ($3.5\times 10^{52}$erg) and the entire gas content of the dSph is pushed to large radii.

For the fiducial disk run, Figure~\ref{fig:disksplash} presents snapshots of the disk density and temperature in the $x-y$ plane (top and bottom rows), and density in the $x-z$ plane (middle row) at 0\,Myr, 730\,Myr and 1.25\,Gyr in the left, middle and right columns, respectively. The snapshots are slices through z=0 (y=0) for the top and bottom (middle) rows and show the 3D density of the particles. Although there is more gas at large radii by the end of the simulation, the density of the disk in the $x-y$ plane remains similar to the initial snapshot with only a small reduction in central density. However the $x-z$ plane slice shows that the final dwarf has evolved into a more spheroidal morphology, particularly for the gas beyond about 0.25\,kpc. Figure~\ref{fig:diskangmom} presents the evolution of the angular momentum over time for the fiducial disk run. The total angular momentum of the gas is represented by the black curves, and the z-component by red curves. Solid curves denote values for all gas particles, which dashed curves relate only to those gas particles within 500\,pc. Although the total z-component of the angular momentum in the dwarf remains almost constant during the run, the z-component within 500pc decreases to less than half of the initial value after 1.2Gyr. While the mid plane density is still relatively high near the centre compared to the initial distribution, the disk is thicker than initially, possibly making it easier for subsequent SN to remove more gas. This final structure, shown in the right panel of Figure~\ref{fig:disksplash}, is typical of all of the disk runs we considered (see Figure~\ref{fig:disk_halo} which presents the corresponding halo evolution). The gas distribution becomes more spheroidal for the lower concentration runs ($c=4$, $r_d=0.2\,$kpc) and the central, higher density regions retain a more disk-like morphology for the higher concentration runs ($c=30$, $r_s=0.05\,$kpc).

The temperature snapshots in the bottom row of Figure~\ref{fig:disksplash} show pockets of warm gas where SNe have recently exploded. Note that the warm gas in this panel is only at $10^4$K as the cooling times are very short at these densities and so it has had time to cool in the interval between the explosion and the snapshot being output. The warm patches in temperature are seen in the corresponding density panels as bubbles where the shock front has pushed gas spherically outwards, leaving low density hot gas around the origin of the explosion.  At the end of the simulation all of the gas is cool ($T< 10^{3.5}$K). 

For comparison we show the same temperature maps for the fiducial spherical run in Figure~\ref{fig:sphere_fiducial_T}. In the fiducial spherical run there is much more warm gas ($10^4$K) in the same region, as the cooling times are much longer. In the final snapshot only around a fifth of the original gas mass within 1\,kpc remains and the temperatures are hotter than for the corresponding disk case. This highlights the differences between the two morphologies. In the case of the disk there is a much higher proportion of gas at higher densities leading to much shorter cooling times. In the spherical case the gas density drops off much more rapidly from the centre, leading to lower densities and longer cooling times. 

The bottom row of Figure~\ref{fig:disksplash} also illustrates the shearing effect of the disk on the gas which has been heated by SNe. For example, the middle panel shows two extended regions in the upper left quadrant of the panel which are left over from earlier SN explosions in regions where the gas density is too low to allow cooling on timescales shorter than the local dynamical times. This effect emphasises the ability of the disk to retain gas that would otherwise be expelled from a non-rotating system. This will also have consequences for the distribution of the metals generated by the SNe - the highly structured nature of the heated gas shows that the distribution of SN products is likely to be very non-uniform. We will explore the impact of this on the metallicity distribution in observed dSphs in a future paper.

The evolution of the gas distributions in the remaining disk runs are shown in Figure~\ref{fig:disk_gas_2}. The top panels show how the evolution changes as a result of varying the initial gas concentration (left), the baryon fraction (middle) and the halo concentration (right); the corresponding enclosed mass profiles are shown in the bottom panels. All panels show the initial profiles as solid lines and the resulting profiles after 1.25Gyr as dashed lines. Figure~\ref{fig:disk_gas_2}(a) shows similar results for varying the gas concentration as in the spherical case. Interestingly, the central density for the most concentrated run (blue curve) is actually slightly higher after 1.25Gyr than at the start of the simulation. This is due to the short gas cooling times in the dense central region of this simulation. As a result, gas which is driven towards the centre by SNe occuring beyond a few 100 pc tends to cool rapidly and contribute to the build-up of a dense, cool central gas concentration.

The disk is able to retain more gas in the most concentrated case than for the corresponding spherical run, with 97 per cent of the original gas mass within 1\,kpc remaining for the disk case, compared to 47 per cent in the spherical case. Increasing the initial concentration of the gas impacts the cooling times in the inner regions of both the spherical and flattened models similarly, and as a result the models with $r_s=0.05\,$kpc and $r_{\rm d}=0.05\,$kpc exhibit similar central densities at the end of the simulation: the mass lost in the spherical case is mostly from radii outside 20\,pc. The ability of the disk to retain significantly more gas overall, is due to the fact that higher density gas in the disk extends to larger radii. We also see a much smaller amount of gas lost from the disk models with initially less concentrated gas ($r_{\rm d}=0.2\,$kpc). In the spherical case almost all the gas is ejected to larger radii with only 7.5 per cent of the original gas remaining within a radius 1\,kpc, compared to 42 per cent for the case of the disk. An initially spherical gas distribution in this case results in the dwarf being destroyed by SN feedback, whereas an initial disk configuration retains its original structure, albeit at a somewhat lower density. 

Figure~\ref{fig:disk_gas_2}(b) shows the effect of reducing the initial baryon fraction of the dSph by a factor of two (blue curves) or four (red curves). Although these two runs have a lower gas mass, the number of particles is kept consistent so that the height of the disk is resolved, resulting in gas particles with masses 5M$_{\odot}$ and 2.5M$_{\odot}$ for $f_b$=0.08 and $f_b$=0.04 respectively. As the Figure shows, the initial reduction in the central gas density leads to a larger amount of gas being ejected from the central regions as expected. A higher fraction of gas is lost within 100pc for those models with a lower initial gas mass, 28 per cent and 30 per cent for $f_b$=0.08 and $f_b$=0.04 respectively, compared to 59 per cent for the fiducial run ($f_b$=0.16). Similarly to the fiducial run, it is the central regions where cooling times are short that are most affected by the SN feedback, as the total mass enclosed within 1kpc is very similar to the initial values.

Figure~\ref{fig:disk_gas_2}(c) shows the results of varying the halo concentration. Note that the central gas densities are slightly different to the fiducial run when a different halo concentration is used. This is due to the method of creating the gas distribution in the disk case, which uses the halo potential to position the gas particles while maintaining hydrostatic equilibrium. Increasing the concentration results in the stellar feedback having no significant effect on the total gas content of the dSph, as 85 per cent of the original gas mass within 1\,kpc is retained. However, the distribution of gas within the dSph is very different to the fiducial run - the gas becomes much more centrally concentrated, similarly to the evolution of the gas distribution in the $r_{\rm d}=0.05\,$kpc model of Figure~\ref{fig:disk_gas_2}(a). Note that in the corresponding spherical run gas is lost from the inner parts of the dwarf, decreasing the density in the centre. However, when the gas is arranged in a disk, mass is only lost from radii $>100\,$pc. In the spherical case only 46 per cent of the original gas mass within 100\,pc remains, whereas 82 per cent remains for the disk. This suggests that gas cooling times play a stronger role in determining whether or not a dSph will retain its gas than modest changes to the depth of the halo potential well, as the initial halo potentials are the same in  both the spherical and disk cases and yet the gas densities evolve very differently. 
Note that the fiducial run ($c=10$) and the $c=30$ run both have the same gas mass within 1\,kpc after 1.25\,Gyr - it is the difference in their central gas densities that will determine the evolution of the dwarf at later times. The gas is much easier to remove from a halo with concentration $c=4$, and by the end of the simulation 90 per cent of the mass within 100\,pc has been removed and the central gas density has dropped by an order of magnitude. 

As the runs with a higher concentration of gas ($r_s=0.05$kpc) and dark matter ($c=30$) retain their high initial central density after 100 SNe, we repeat the same runs with 200 and 500 SN events. The resulting gas and halo density profiles are shown in Figure~\ref{fig:disk_concentration}. When comparing the higher concentration run for the sphere and the disk (Figure~\ref{fig:sphere_gas} (b) and Figure~\ref{fig:disk_gas_2} (c)) it seems that gas cooling time is more important for retaining gas than the depth of the potential well, as the spherical case results in gas removal from the centre of the dwarf in contrast to the corresponding disk simulations. However, if we consider the spherical and disk models separately, and look at the effect of gas or halo concentration on the ability of a halo to retain gas in the presence of larger numbers of SNe, we see a different behaviour for each initial morphology.

Following the explosion of 500\,SNe, the spherical model loses all its gas even in the cases of higher concentrations of either the gas or dark matter. The gas density is reduced to less than $10^4$M$_{\odot}$kpc$^{-3}$ in both cases (i.e. below the limits of the plot). After 200 SN events, the spherical run retains more gas in the higher gas concentration case where the gas particles have shorter cooling times, resulting in the final central density being potentially high enough to host more star formation ($2.4-7.4\times10^9$M$_{\odot}$kpc$^{-3}$~\citep[]{2012MNRAS.421.3464P, 2016MNRAS.tmp..504R}). This is due to both the binding energy in the inner 60pc being slightly higher ($1.95\times 10^{52}$erg and $1.39\times 10^{52}$erg for $r_s=0.05kpc$ and $c=30$ respectively) and the initial central gas density being higher, resulting in shorter gas cooling times allowing for energy to be radiated away. 

When the morphology is initially spherical, a change in the gas density has a larger impact on the cooling times than for the corresponding disk models (where the cooling times are already short) and hence changing the gas density in the spherical case results in a significant change to the overall evolution. In this case, a higher initial gas density is more effective at retaining gas than the deeper potential well due to a more concentrated dark matter halo.

For the disk morphology, 200 SN events have a broadly similar impact to that due to 100 SN and the central gas densities remain very close to the initial values. However, larger difference are seen after 500 SNe. The $c=30$ run retains significantly more gas than the $r_s=0.05kpc$ run after 500 SN events, with a final central gas density of $10^{8}M_{sun}kpc^{-3}$. Here the depth of the halo potential is keeping the gas bound to the dSph halo. The total binding energy of the gas is roughly the same as the injected SN energy ($5.2\times10^{52}$ for $r_s=0.05kpc$ and $4.66\times 10^{52}$ for $c=30$). In contrast to the spherical case, for the disk the higher halo concentration is more effective at retaining gas, again potentially resulting in a qualitatively different evolutionary path for the dwarf. In the case where the disk and the sphere have the same potential energy, the cooling time of the gas determines whether the dwarf will be able to retain enough gas to continue forming stars (as in the disk model) or if a large fraction of gas will be ejected (as in the spherical model). If the cooling time is long as in the spherical case, making the potential well deeper does not improve the chances of the dwarf retaining gas. However, in models for which the gas cooling time is already shorter than the dynamical time, a deeper potential well increases the fraction of dense gas which is retained. 

The time evolution of the fractions of gas particles with velocities greater than that formally required to escape the potential for a selection of disk runs is shown in Figure~\ref{fig:vesc_disk}. The solid curves show the fraction of gas located within $r<1\,$kpc which have velocities above the escape velocity; the dashed lines show the same for gas at $1\,$kpc$<r< 5\,$kpc. As expected, Figure~\ref{fig:vesc_disk}(a) shows that as more SN energy is injected, more gas is driven above the escape velocity. In the three simulations with varying numbers of SN events, all the gas remaining within 1\,kpc is below the escape velocity by the end of the run and so will stay bound to the dwarf. The fraction of gas at $r > 1$\,kpc above the escape velocity declines towards the end of the simulation, but this gas is unlikely to fall back into the central regions due to the infall time required (the average free fall time for gas at $1<r<5$ kpc is 1Gyr). This suggests that although only a small fraction of gas is removed from the main disk, SN feedback is likely very important for keeping the gas hot and extended, allowing it to be removed easily by other external processes (e.g. ram pressure stripping) when the dSph falls into a host halo at lower redshifts. The same effect is seen in the simulations which vary the other model parameters. Figures~\ref{fig:vesc_disk}(b) and (d) show results for varying the gas and halo concentration, respectively. 
In both cases, the fact that the most concentrated runs give a larger fraction of gas at the escape velocity at larger radii is simply due to the fact there is less gas at these radii. Only the small mass in high velocity gas is able to reach these radii and hence a higher fraction is formally unbound. Again the fraction decreases to a negligible amount over time and this will likely stay bound to the dwarf in the absence of external perturbations.

Figure~\ref{fig:disk_halo} shows the effect on halo density profile evolution of varying (a) the number of SN events, (b) the gas concentration, (c) the baryon fraction and (d) the halo concentration, for the disk case. Increasing the number of SN events (and hence the energy injected) decreases the central density of the halo as more gas is pushed out to large radii. This is a small decrease in comparison to the spherical case in which 100 SN events cause an order of magnitude decrease in the central halo density due to the sudden removal of a large amount of gas. Changing the gas concentration or the baryon fraction has no effect on the density of the halo (panels b and c respectively), due to the small amount of gas removed in comparison to the spherical case. 

The Figure clearly shows that the only parameter which has a significant impact on the evolution of the halo density profile of the halo is the initial halo concentration (Figure~\ref{fig:disk_halo}~(d)). In the simulation with halo $c=4$, the total potential inside $60\,$pc is initially $52$ per cent lower than in the $c=10$ run. In addition, the gas contributes 39 per cent to the total potential inside $60\,$pc for the $c=4$ run compared to 32 per cent in the $c=10$ case. These two effects both serve to increase the impact on the dark matter profile of gas removal. The shallower potential well in the $c=4$ run allows more gas to escape, which has a proportionately larger effect on the potential well as a whole since the gas contributes a larger fraction of the potential. This is the only run in the disk case for which the final halo density profile is significantly shallower than the initial profile. All other runs with 100 SNe do not remove enough gas to have any major impact on the halo density; while the runs with 200 and 500 SNe (panel (a)) reduce the central density by a factor of a few, the inner profile retains its initial cusp.
\subsection{Metallicity}
\label{sec:metallicity}
\begin{figure}
\includegraphics[width=0.5\textwidth]{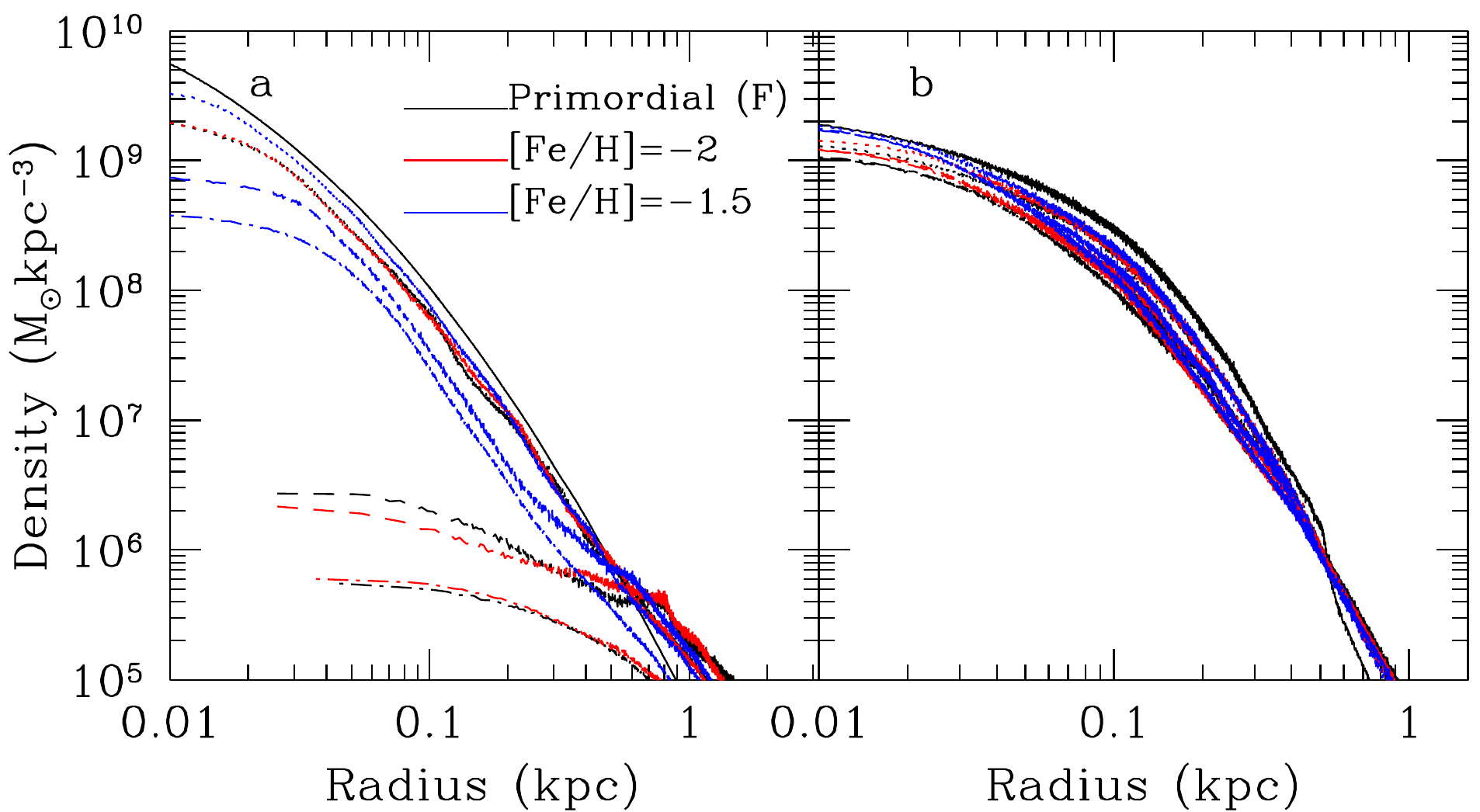}\vspace*{-3mm}
\caption{Evolution of the gas density for simulated dSphs assuming varying gas metallicities: primordial (our fiducial run, black), [Fe/H]=-2 (red) and [Fe/H]=-3 (blue) for a spherical (left) and disk (right) morphology. The different curves represent the initial profile (solid) and the density profile after 0.4Gyr (dotted), 0.8Gyr (dashed) and 1.2Gyr (dot-dashed) of evolution.}
\label{fig:metallicity} 
\end{figure}

In our standard runs we assume gas of primordial composition and do not model the effects of metals produced by supernovae. We repeat the standard runs assuming a metallicity for the gas of [Fe/H]$=-1.5$ and $-2$, using approximate cooling functions based on those described in~\citet{1993ApJS...88..253S} down to $10^{4}$K. Radiative cooling below $10^{4}$K is included as described in~\citet{2008Sci...319..174M} from the de-excitation of fine structure and metastable lines of C, N, O, Fe, S and Si. The resulting gas density profiles for the spherical and disk fiducial runs are shown in panels a and b of Figure~\ref{fig:metallicity}, respectively. The initial density profile is the same for all metallicities and is represented by the black solid line, other lines show the profile at 0.4Gyr (dotted), 0.8Gyr (dashed) and 1.2Gyr (dot-dashed). The radial density profile of the primordial run over time is shown as black lines, and as red (blue) lines for [Fe/H]=-2 (-1.5). 
 
For the spherical case, [Fe/H]=- 2 (red) alters the final density profile by a small fraction, leaving the result unchanged. However, [Fe/H]=-1.5 (blue) shows a much larger difference, with the central density profile only reduced by less than 2 orders of magnitude, compared to four orders of magnitude for the other two cases. The difference between the runs at earlier times (0.4Gyr) is much smaller as the gas density is still relatively high resulting in short cooling times for all runs regardless of the metallicity. As the density is reduced by the continuing SN events the role of metallicity becomes more important in determining the cooling time of the gas.
The difference in the resulting density profiles for the disk case is much smaller, as the cooling times in the disk are already shorter than in the spherical case due to there being a larger fraction of gas at higher densities. Even in the highest metallicity run ([Fe/H]=-1.5, shown in blue) the inclusion of metal cooling has very little effect on the central density, showing that here it is the density that is the most important term in calculating the cooling time.

\section{Discussion}
\label{sec:discussion}

The simulation results presented in the previous section show that the most likely outcome of star formation in a dark matter halo of mass $3\times10^7M_{\odot}$ is the total expulsion of most of the remaining gas from the system due to SNa feedback. The precise fraction of gas that remains within 1kpc of the halo at the end of our simulations is a strong function of the gas cooling time in the regions where the SNa explosions occur. The more rapid gas cooling in regions of high gas density has a greater impact on the ability of a halo to retain gas than either the depth of the potential well or the initial morphology of the gas distribution. However, as our disk models have larger fractions of gas at higher densities than the spherical distributions, all the disk models retain significantly more gas than the corresponding spherical models in our simulations. We also note that for a disk-like gas distribution, the path of least resistance for outflowing gas is one that is perpendicular to the plane of the disk~\citep[see, e.g.,][]{2011MNRAS.415.1051B}, allowing more gas to be retained in the plane of the disk, especially in the central regions. In most of the spherical models, gas is easily expelled at all radii including the central regions. The exception to this is for the spherical model with the highest gas concentration ($r_{\rm s}=0.05$kpc) which maintains a high gas density in the centre, although gas is still lost from larger radii. 

The importance of cooling time is confirmed by the impact on our results of increasing the initial gas metallicity. A global increase of the initial metallicity to [Fe/H]=-1.5 significantly enhances the ability of a spherical gas distribution to maintain gas at high density in the inner regions. Even for dSphs with initial gas disks, Figure~\ref{fig:metallicity}(b) shows that increasing the metallicity leads to almost a doubling of the gas density within $20$pc at the end of the simulation relative to a simulation assuming primordial abundances. 

However, pre-enrichment of the gas to metallicities as high as [Fe/H]$\sim$-1.5 in dSph progenitors is not favoured by observations. High-resolution spectroscopic studies of the stellar populations of dSphs like Ursa Minor and Draco show that they were continuously forming stars for $\sim$3 Gyr from z\,$\sim$\,12 to z\,$\sim$\,1~\citep[e.g.][]{Weisz2014}. In particular, the star formation history of Ursa Minor suggests that roughly 10 percent of the stars were formed by z\,$\sim$\,10 while almost forty percent of the stars were in place by z\,$\sim$\,4~\citep[][ their Figure 1]{Weisz2014}. Detailed modelling of the metallicity distribution function (MDF) of Ursa Minor favours models which include both gas infall~\citep{2013ApJ...779..102K} and outflows~\citep{2015MNRAS.449..761U}. Further, the tentative age-metallicity relation for stars in Ursa Minor obtained by~\cite{Cohen2010} suggests that the more metal-rich tail of the metallicity distribution is populated by stars which formed more recently than the main stellar population. The inflow of gas with primordial abundances, with star formation potentially being triggered by mergers with other, similar mass haloes, could therefore explain both the MDF and the age-metallicity relation. Once the halo reached a larger mass it was able to retain enriched gas and form stars with higher [Fe/H]. A model in which the halo continued to form stars at a low level for sufficient time after growing its potential to give rise to stars with enhanced [Fe/H] and reduced $\alpha$/Fe ratios appears to be consistent with all the observed data. For example, the data in both~\cite{2015MNRAS.449..761U} and~\cite{Cohen2010} suggest that only the youngest (and most metal-rich) stars in Ursa Minor exhibit reduced $\alpha$/Fe.

Although a globally raised metallicity is unlikely to be the origin of the short gas cooling times required to facilitate gas retention, the higher gas densities typical in flattened morphologies remain a plausible explanation. If the gas initially has a spherical distribution, we have shown that either the dark matter must be more concentrated than is expected for typical haloes of this mass at z\,$\sim$\,10~\citep{2015MNRAS.452.1217C}, or gas cooling must have generated a gas distribution with a scale-length roughly an order of magnitude smaller than that of the dark matter, implying significantly stronger collapse than is seen in typical haloes at this mass scale~\citep[e.g.][]{2016MNRAS.tmp..504R}.  Even in this case, the star formation would be limited to the central regions of the halo (i.e. within a few tens of pc), and a subsequent evolutionary process would be required to disperse these stars over a larger fraction of the halo. The high stellar and dark matter densities in the inner regions would mean that externally-driven processes such as the tidal sculpting usually invoked to change the morphology of dSph stellar distributions would have little effect~\citep[e.g.][]{2006MNRAS.367..387R,Penarrubia2010}. Similarly, mergers with haloes of comparable or larger masses (and hence of lower density) would be unable to inflate the spatial scale of the stellar distribution by an order of magnitude and would be more likely to result in the formation of a nucleated dwarf galaxy~\citep[e.g.][]{2015ApJ...812L..14B,2013MNRAS.432..274A}.

Obviously, the presence of dense gas with short cooling times is a pre-requisite for further star formation. Given that only a small fraction of the stars in the less luminous dSphs (e.g. Ursa Minor, Draco and Sextans) have metallicities as high as [Fe/H] = -1.5 at the present epoch~\citep{2013ApJ...779..102K}, we infer that the primary determinant of whether a dSph will exhibit extended star formation is the overall fraction of gas which is at high density when the first stars form. Although this gas may not yet have formed stars, it is only this gas which will be retained by the dSph. In other words, it is not sufficient for cooling in a dSph progenitor to achieve a high gas density solely at the centre of the halo  - it must have high density gas distributed throughout the volume which will ultimately be occupied by the stellar population of the dSph. Given that a disk morphology is likely to have a larger fraction of high density gas, in particular outside the central few tens of parsecs, our simulations are consistent with the idea that the progenitors of the Milky Way dSph satellites experienced the bulk of their star formation at an epoch when their baryonic components had a disk morphology. The transformation of rotation-supported disks into pressure-supported spheroids through the action of external gravitational tides has already been discussed by a number of authors in the literature~\citep{2001ApJ...547L.123M,2001ApJ...559..754M,2009MNRAS.397.2015K,2012ApJ...751...61L,2016ApJ...818..193T}. ~\cite{Nayakshin2013} have discussed the role that AGN outflows from the Milky Way could have played in that process.

In our fiducial, spherical model the gas is spread out over a sphere of radius $\sim20$kpc by the end of the 1.25Gyr simulation. The gas at large radii ($>5$kpc) is at low enough densities to be stripped on infall to the MW and would become a part of the hot halo which is likely to have a density of few $\times10^2$\,M$_\odot$\,kpc$^{-3}$ at a radius of 100\,kpc from the Galactic Centre~\citep{2006MNRAS.369.1021M}. The gas at radii $1$kpc\,$<r<5$kpc could possibly fall back, although at the end of our simulations its low density means that its cooling time is very long. Observational determinations of the star formation histories of dSphs are not yet sufficiently precise to distinguish continuous star formation from bursty star formation punctuated by periods of $\sim$100\,Myr during which gas which remained in the halo could potentially cool and re-ignite star formation~\citep[see e.g.][]{Weisz2014}. However, if our simulated dwarf is a progenitor for a dSph like Draco or Ursa Minor we expect further star formation and outflows, which would tend to suppress the cooling and infall of the gas at these radii and we therefore need to consider further sources of low-metallicity gas.

In all our simulations, the gas distributions were initially smooth although the injection of SN energy rapidly leads to a highly structured ISM. In a future paper, we will investigate the impact of an initially clumpy ISM, such as might be produced by stellar winds prior to the onset of SN feedback~\citep{2016MNRAS.456L..20B}. Dense gas clumps are more resilient to feedback~\citep{2016MNRAS.456L..20B} and could therefore increase the amount of gas retained by the halo. In the context of AGN feedback impacting on the gas at the centre of a galaxy, \cite{2014MNRAS.441.3055B} has shown that significantly more gas remains after the passage of an AGN outflow through a non-uniformly distributed ISM compared to a smooth distribution. It is therefore reasonable to expect that similar behaviour may be seen in the context of a dSph progenitor.

In this paper, we evolved our dSph progenitors in isolation. As a result, we have neglected a number of external effects in our simulations which could potentially impact on our conclusions, namely gas accretion and re-ionisation. The more important of these at this redshift is likely to be gas accretion. \cite{2013ApJ...779..102K} find that a gas accretion model is a good fit to the narrow metallicity distribution functions for Ursa Minor and Draco (and, indeed, most other dSphs). Those authors find that it is unlikely that the stars which formed later in these dwarfs were composed of gas enriched by the first SN, but rather they formed from more pristine gas subsequently accreted onto the halo. In this case retention of the gas initially present in the dSph may not be required for further star formation. If additional gas is continually accreted by the halo as star formation proceeds, then the requirement is actually that the outflows generated by the on-going star formation are unable to  prevent the accretion and cooling of this gas. Simulations by~\cite{2015MNRAS.452.1026L} have shown that this infall may have been filamentary: the low filling factor of these filaments would naturally increase the likelihood of gas being able to reach the central few 100 pc of the halo even in the presence of a SN-driven outflow from the dSph. Indeed~\citet{2015MNRAS.451.2900K} find that SN-driven outflows have very little influence (irrespective of the strength) on the rate of inflowing gas, showing that cold filamentary accretion at high redshift allows for the deposition of low-angular momentum gas in the central regions, enhancing the gas density. 

Given that our simulations cover the period from of z\,$\sim$\,10 to z\,$\sim$\,4, it is likely that the reionization of the Universe took place during this time~\citep[reionization began between z\,$\sim$\,10-15 and was complete by z\,$\sim$\,6; see e.g.][and references therein]{Planck2016}. It has been suggested that this could have contributed to the quenching of star formation in low mass halos~\citep[]{2016MNRAS.456...85S}. Observationally, however, Ursa minor and Draco show evidence for extended star formation during this time period and beyond and no local group dSph exhibits a feature in its star formation history which is unambiguously associated with reionisation~\citep{Weisz2014}. While this suggests that it is unlikely that reionization was responsible for the cessation of star formation ~\citep[though see][for a discussion of possible self-shielding of gas in dSphs during re-ionisation]{2010MNRAS.402.1599S}, it could have played a role in limiting the supply of cold gas from accretion. In this way, it could possibly account for differences between the dSphs and the ultra faint dwarfs: those haloes whose star formation was truncated by reionisation became the ultra faints, while those that survived reionization and continued their forming stars until their supply of gas was exhausted became ``classical'' dSphs.

In agreement with previous work~\citep[e.g.][]{2010Natur.463..203G,2005MNRAS.356..107R,1996MNRAS.283L..72N} we find that it is possible to form dark matter cores providing halo concentrations are lower at high redshift, and enough gas is ejected from the central regions. For the removal of gas to have an effect on the halo the gas in the centre must make a significant contribution to the potential and/or a significant amount of mass must be removed at once. However, as noted by~\cite{Penarrubia2010}, if a core is formed early in the evolution a dSph satellite, it is questionable whether it could subsequently survive to the present day without being tidally disrupted by the Milky Way.

Independent of our other model parameters, we find that the SN rate in our simulations must be below 0.5 Myr$^{-1}$ to allow dense gas to remain in the dwarf at the end of the simulation. As discussed in section~\ref{sec:simdesc}, the stellar mass of the present-day Ursa Minor would imply a rate of at least twice this (assuming a constant star formation rate). There are several possible explanations for this. First, we have assumed that all the star formation occurred in a single halo - it is possible that some of the low-metallicity stars in Ursa Minor formed in a separate halo which later merged with the main halo. Given that the halo mass has increased by at least an order of magnitude since z\,$\sim$\,10, it is possible that at least one other halo that contributed to the dark matter content of this dSph also contained stars and therefore each individual halo could have hosted a smaller number of SNe. Secondly, we note that the star formation history for Ursa Minor presented in~\cite{Weisz2014} implies that approximately 10 per cent of the stars in Ursa Minor were in place by z\,$\sim$\,10. This would correspond to a stellar mass of $1-2\times10^{4}$\,M$_\odot$ which would result in 100-200 SNa explosions, comparable to the numbers in those of our simulations which resulted in gas retention. Thirdly, in estimating the expected numbers of SNe we have assumed a~\cite{1955ApJ...121..161S} initial mass function for the stellar population. While the stellar mass function for low-mass stars in Ursa Minor has been shown to be identical to that of the globular clusters M15 and M92~\citep{2002NewA....7..395W}, the high-mass end of the mass function in an old stellar population can only be inferred indirectly from chemical signatures. Thus, it is possible that the high-mass stellar IMF in high-redshift dSphs was significantly different from that seen in star-forming regions at z\,$\sim$\,0.

\section{Conclusions}
\label{sec:concl}
In this paper, we addressed the issue of whether a typical $z=10$ dSph progenitor could maintain an extended burst of star formation having initially acquired a gas fraction comparable to the universal baryon fraction. In contrast to many previous studies, our $z=10$ haloes had masses of $3\times10^7$M$_{\odot}$, appropriate for the $z=10$ progenitors of haloes with masses of $\sim10^9$M$_{\odot}$ at $z=0$. We included the energy injected by SN explosions, and followed the evolution of the gas distributions for $1.25$Gyr. Our primary conclusion is that gas retention by haloes of this mass is very challenging and requires halo concentrations or initial gas distributions which are atypical of haloes at $z=10$. 

We varied several parameters to create our model dwarf galaxies including the morphology of the initial gas gas distribution, the number of SNe, the gas concentration, the baryon fraction and the halo concentration. We find that the cooling time of the gas in the central regions is the most important factor to retain sufficiently high central densities to allow star formation to continue. For both spherical and disk gas morphologies increasing the gas concentration increases the central density, resulting in the gas having shorter cooling times which reduces the impact of the SN feedback on the ISM. A flattened gas distribution (disk) is overall more efficient at retaining gas than a spherical distribution due to the fact that the gas densities are generally higher initially. However, even with a high gas or halo concentration, we find that high density gas only remains at the end of the simulation period if the number of SN events is lower than that expected from a Salpeter IMF (assuming a total stellar mass of $\sim 3\times10^5$M$_{\odot}$). 

Our finding that the haloes of dSph progenitors at $z=10$ were not typical of haloes at that redshift, combined with the data on the metallicity distributions of observed Milky Way dSph satellites, suggests that the morphology of infalling gas may be a key factor which determines the ability of a given halo to support extended star formation. Successful dSphs may therefore result from progenitors which were outliers in the distribution of halo/gas properties at $z=10$, as well as experiencing gas accretion at an appropriate rate and with an appropriate morphology so that SN-driven outflows were unable to prevent further gas accretion. This would naturally explain the apparently non-linear mapping required for the abundance matching of haloes from cosmological simulations with observed galaxies, which gives rise to the ``Too Big To Fail'' problem and the apparent stochasticity of galaxy formation at the faint end of the galaxy luminosity function.

In future work, we will investigate the extent to which a clumpy initial gas distribution might affect the ability of a dSph to retain gas following an initial burst of star formation. We will also explore the impact of external AGN outflows on the evolutionary history of the dSphs as discussed in~\citet{Nayakshin2013}.

\endgroup

\section*{Acknowledgements}
The authors would like to thank Sergei Nayakshin, Walter Dehnen and Hossam Aly for useful discussions. We thank the anonymous referee for their comments which improved the final version of this paper. CRC and MAB were supported by a Science and Technology facilities council (STFC) PhD studentship. CP acknowledges support from the Australian Research Council (ARC) Future Fellowship FT130100041 and Discovery projects DP130100117 and DP140100198. Figures~\ref{fig:disksplash} and~\ref{fig:sphere_fiducial_T} were produced using SPLASH~\citep{Price_2013}. This work used the DiRAC Complexity system, operated by the University of Leicester IT Services, which forms part of the STFC DiRAC HPC Facility (www.dirac.ac.uk ). This equipment is funded by BIS National E-Infrastructure capital grant ST/K000373/1 and  STFC DiRAC Operations grant ST/K0003259/1. DiRAC is part of the National E-Infrastructure.




\bibliographystyle{mnras}
\bibliography{ref}



\appendix
\begingroup
\let\clearpage\relax
\section{Stability test}
\label{sec:appstab}
\begin{figure}
\includegraphics[width=0.5\textwidth]{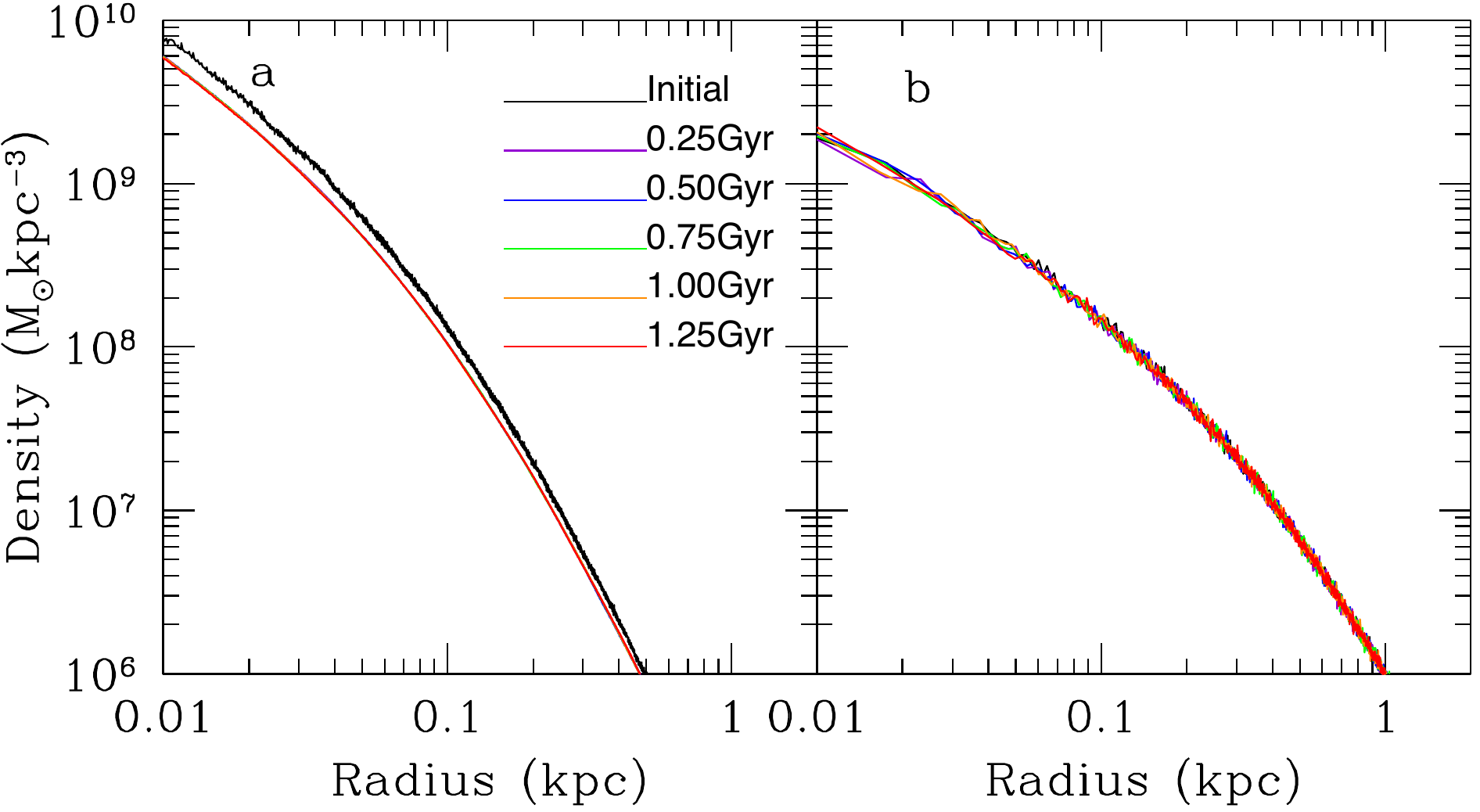}
\caption{Evolution of the \textbf{(a)} gas and \textbf{(b)} halo density profiles over time for gas set up in a spherical distribution with no feedback.}
\label{fig:sphere_stability} 
\end{figure}
\begin{figure}
\includegraphics[width=0.5\textwidth]{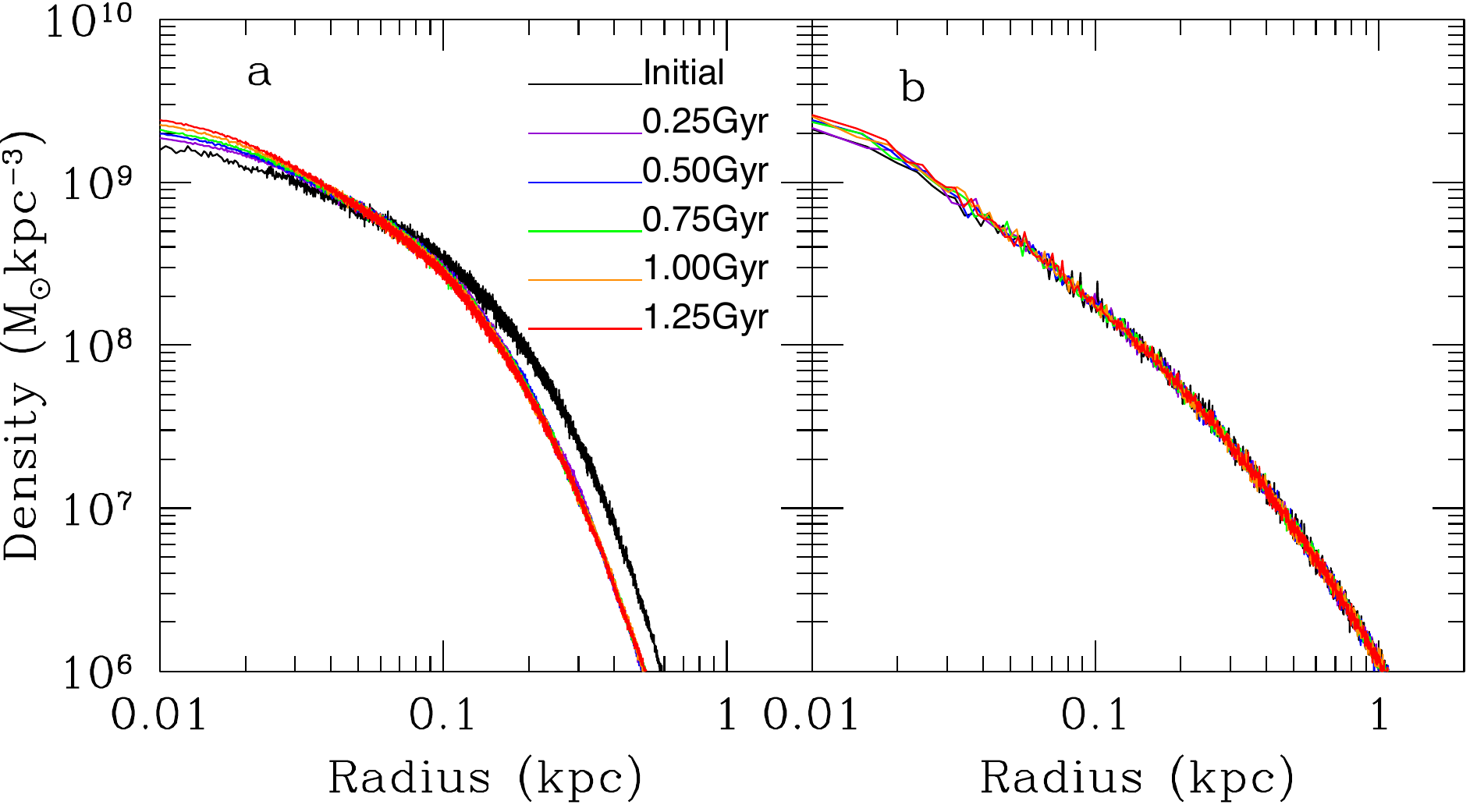}
\caption{Evolution of the gas (a) and halo (b) density profiles over time for gas distributed in a disk with no feedback.}
\label{fig:diskstab} 
\end{figure}

In this Appendix, we consider the long term evolution of both the spherical and disk gas distributions in order to test the stability of our initial conditions. The initial conditions were set up as in section~\ref{sec:simdesc}, and the simulations were run without cooling or SN feedback for 1.25Gyr. The density profiles for the gas and halo in each distribution are shown in figures~\ref{fig:sphere_stability} and~\ref{fig:diskstab} for the sphere and disk respectively. There is some initial settling to an equilibrium configuration due to the fact that the gas and halo components are set up separately, this is complete after $\sim$250Myr, so for all the runs in the paper we allow the initial conditions to settle for 300Myr before switching on cooling and feedback.

\section{Convergence test}
\label{sec:appconvergence}
\begin{figure}
\includegraphics[width=0.5\textwidth]{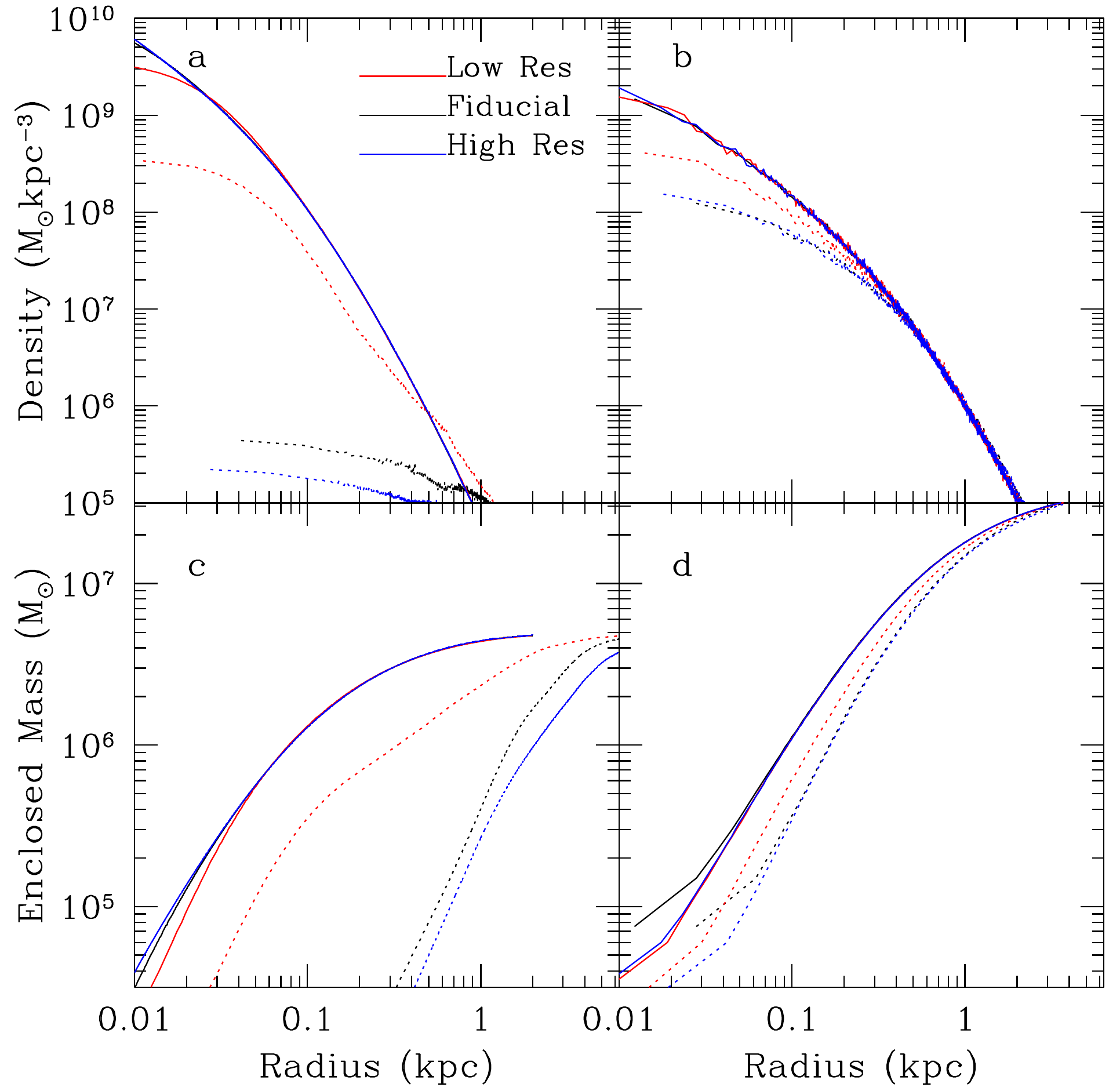}
\caption{Evolution of the density (\textbf{top})  and enclosed mass (\textbf{bottom}) profiles for gas (left) and dark matter (right) for simulated dSphs with an initially spherical gas distribution. In all panels, the solid curves show the initial profiles, and the blue, black and red dashed lines show the resulting profiles after $1.25$\,Gyr of evolution for simulations with gas particles masses of $1$\,M$_{\odot}$, $10$\,M$_{\odot}$ and $100$\,M$_{\odot}$,  respectively.}
\label{fig:sphere_res} 
\end{figure}

\begin{figure}
\includegraphics[width=0.5\textwidth]{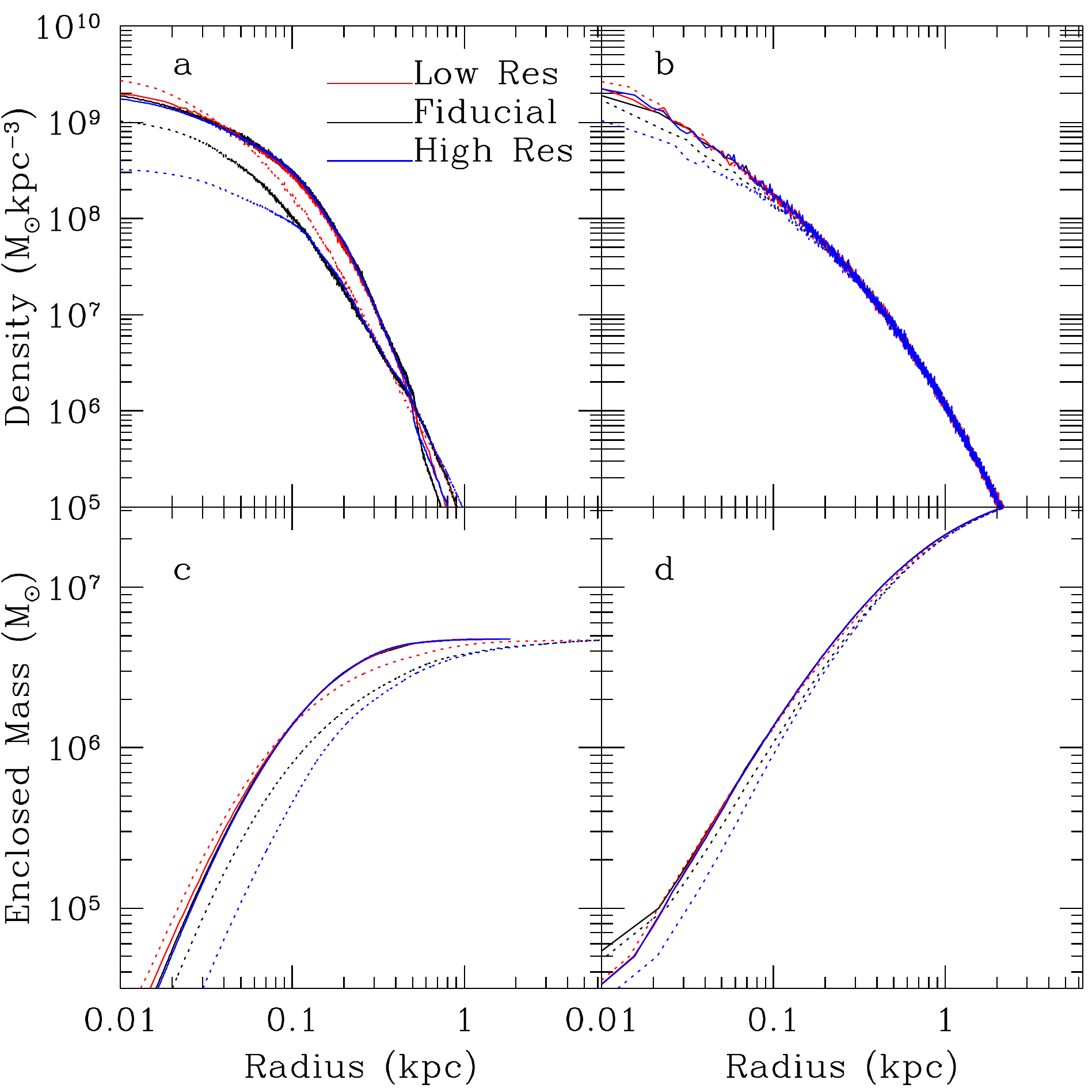}
\caption{Evolution of the density (\textbf{top})  and enclosed mass (\textbf{bottom}) profiles for gas (left) and dark matter (right) for simulated dSphs with gas initially distributed in a disk. In all panels, the solid curves show the initial profiles, and the blue, black and red dashed lines show the resulting profiles after $1.25$\,Gyr of evolution for simulations with gas particles masses of $1$M$_{\odot}$, $10$M$_{\odot}$ and $100$M$_{\odot}$, respectively.}
\label{fig:disk_res} 
\end{figure}

To explore the impact of mass resolution on our conclusions , we re-run the fiducial simulation (see table 1) at mass resolutions ten times higher and ten times lower for both the sphere and the disk. We keep the mass of gas into which the SN energy is injected constant ($1000$\,M$_{\odot}$) by changing the \textit{number} of particles receiving energy according to the new mass resolution. Figures~\ref{fig:sphere_res} and ~\ref{fig:disk_res} show the evolution of the profiles for the gas and halo densities in the spherical and disk cases, respectively. 

In the spherical case, reducing the mass resolution to $100$\,M$_\odot$ significantly increases the mass of gas which remains after 1.25 Gyr. This emphasises the need for high resolution simulations to capture SNa feedback in dSph progenitors in a realistic way. We note that with $100$\,M$_\odot$ resolution, only 10 neighbouring particles receive energy from a SNa which results in too much of the energy being radiated away before it can be deposited in the remaining gas. As is well-known from previous simulations, this means that at low resolution, various numerical tricks (e.g. delayed cooling etc) are required to mimic the impact of SNa feedback. 

It is reassuring to note, however, that increasing the resolution by a factor 10 relative to our fiducial run has a limited impact on the evolution of the dSph in the spherical case. In particular, the SNa feedback moves almost all the gas to beyond $1$\,kpc in both the fiducial and higher resolution runs.

In the case of a dSph model with a gas disk, increasing the resolution has an impact on the gas distribution in the inner $100$\,pc. This is because the higher resolution model has longer gas cooling times in the inner regions, and the SNe are therefore more effective at removing gas from the inner parts of the dSph. However, Figure~\ref{fig:disk_res} shows that beyond $\sim100$\,pc the profiles of both gas and dark matter are indistinguishable between the fiducial and high resolution runs. The inner 100pc contain less than 10 per cent of the gas mass in the dSph and therefore we consider that we have correctly captured the evolution of the bulk of the gas in the dSph.

More importantly, we note that our conclusions based on simulations at our fiducial resolution are conservative in the sense that, if anything, we have underestimated the mass of gas that SNa feedback will remove. Our conclusion that it is difficult for a low-mass dSph progenitor halo to retain the gas needed to sustain extended star formation, and hence that the observed dSphs must have been unusual in their initial properties (gas morphology and/or concentration, halo concentration), remains valid at higher resolution.



\section{Feedback implementations}
\label{sec:appfeedback}
\begin{figure}
\includegraphics[width=0.5\textwidth]{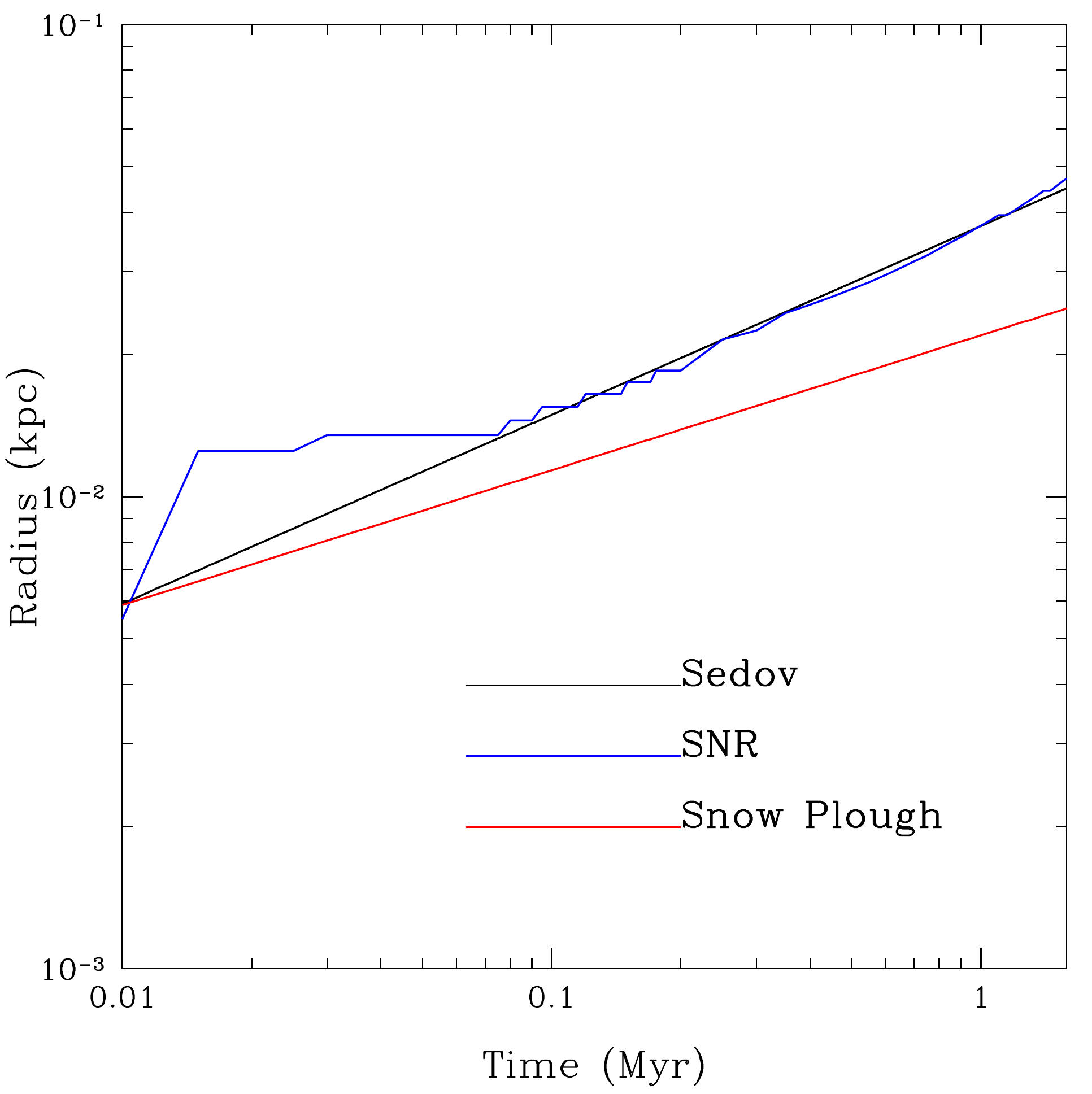}
\caption{Evolution of the radius of the SNa-driven shell with time where the shell-radius is defined by the peak of the radial density distribution around the SNa. The dense shell forms at ~0.04Myr, and subsequently follows the Sedov-Taylor solution. The black line shows the Sedov-Taylor solution for the growth of the shock front resulting from the injection of $10^{50}$\,ergs of purely thermal energy into a medium of uniform density $5\times10^{8}$\,M$_{\odot}$\,kpc$^{-3}$. The red line shows the time evolution for the momentum-driven phase of shell growth. At later times, the evolution of the SN remnant would follow this line, in the absence of external perturbations from other SNe.}
\label{fig:shellrad}
\end{figure}

\begin{figure}
\includegraphics[width=0.5\textwidth]{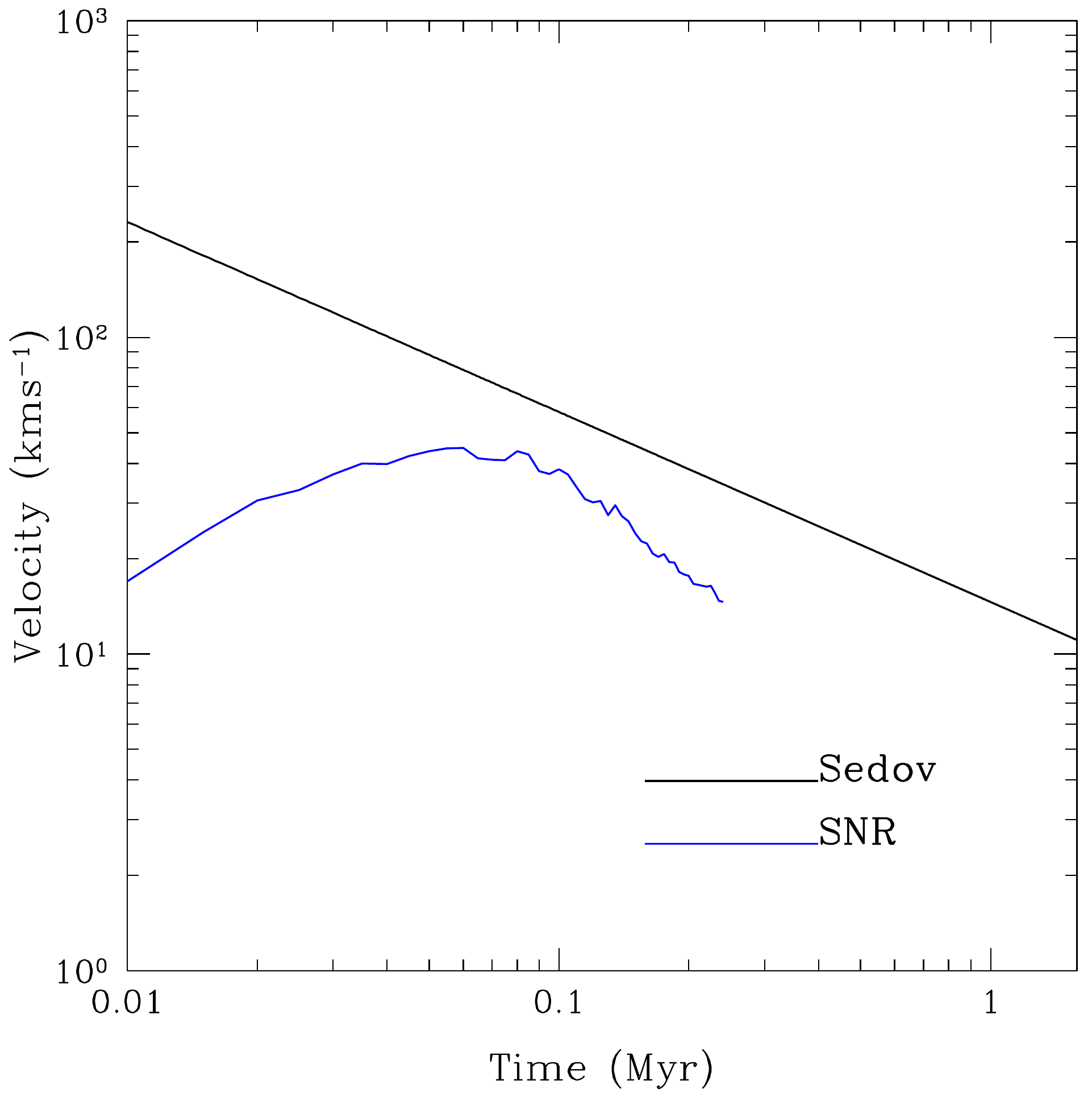}
\caption{Evolution of the radial velocity of the SNa-driven shell with time where the shell velocity is defined by taking the peak of the average radial velocity in radial bins of width 1pc (blue). The black line shows the Sedov-Taylor solution for the velocity of the shell resulting from the injection of $10^{50}$\,ergs of purely thermal energy into a medium of uniform density $5\times10^{8}$\,M$_{\odot}$\,kpc$^{-3}$.}
\label{fig:shellvel}
\end{figure}

\begin{figure}
\includegraphics[width=0.5\textwidth]{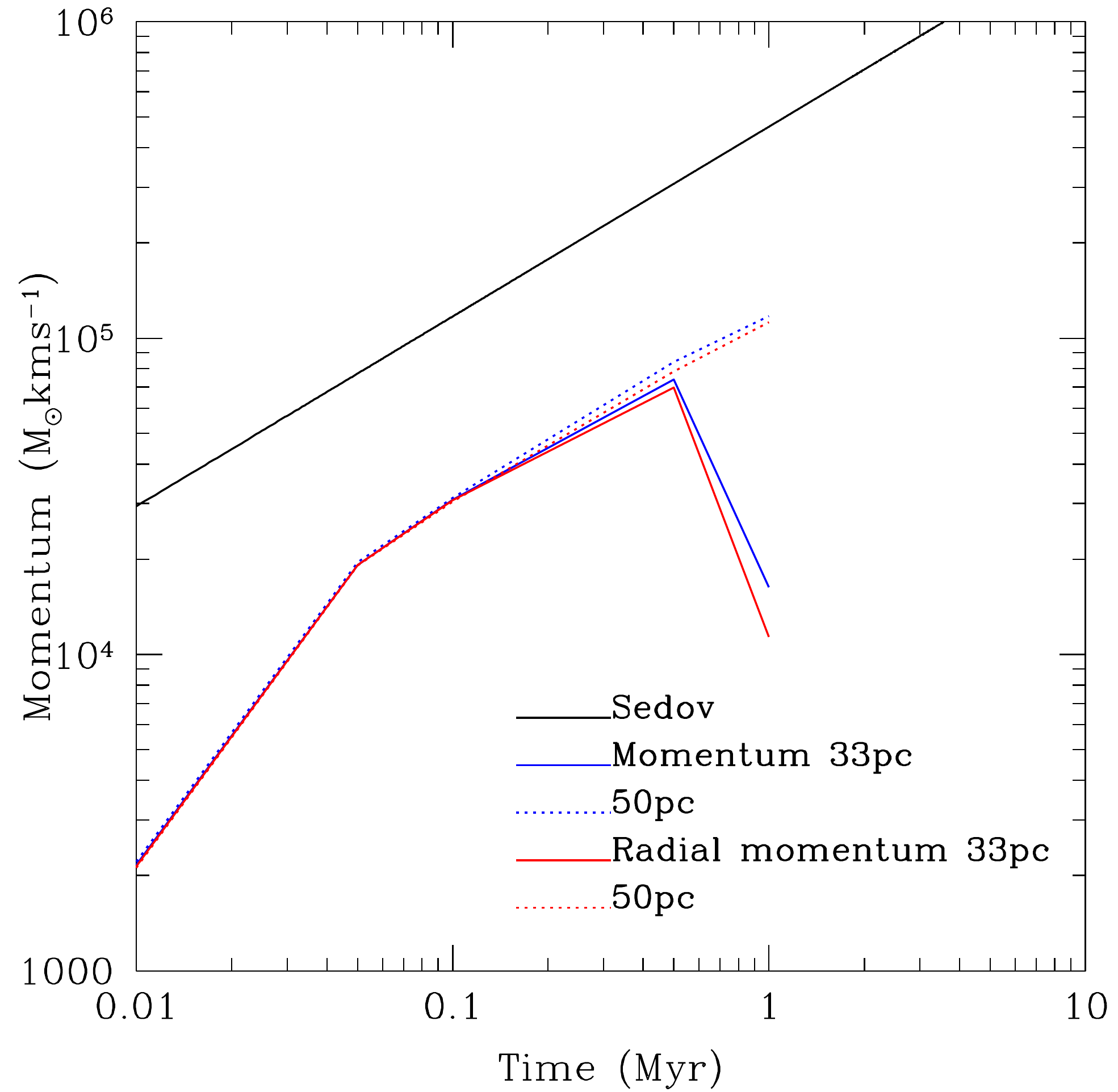}
\caption{Evolution of the total (blue - magnitude) and radial (red) momentum in a box of 33pc (solid lines) and 50pc (dashed lines) centred on the SN. The Sedov solution is shown by the solid black line.}
\label{fig:boxmom}
\end{figure}

Modelling the evolution of a SNR is computationally demanding due to the large dynamic range needed to capture physical evolution on a wide range of length scales. Large scale cosmological simulations lack the resolution needed to resolve each SN individually due to numerical over-cooling~\citep[see, e.g,][]{2012MNRAS.426..140D} as the high-mass gas particles result in most of the injected energy being radiated away before it can impact the ISM . 

In this appendix, we show that we can model individual SNa events with a mass resolution of $10$\,M$_{\odot}$ by injecting the nearest 100 gas particles with $10^{50}$erg of thermal energy, kernel-weighted according to their distance from the star particle. At a density of $5\times10^{8}$\,M$_{\odot}\mathrm{kpc}^{-3}$, this corresponds to a spherical radius of $\sim 70$pc. 

Although we are unable to resolve the initial 'free expansion' phase of a SNR (when the ejecta expand freely while the mass swept up by the forward shock is smaller than the mass ejected) due to the number of particles that are given energy, we can resolve the Sedov Taylor phase. In this phase, an adiabatic blast wave expands into the ISM with $r_{\rm shell} \propto t^{2/5}$, and total energy is conserved, consisting of 73 per cent thermal and 27 per cent kinetic energy~\citep{2012MNRAS.419..465D}. The movement of the shell is still pressure-driven while the gas interior is hot. When the cooling time of the gas in the shell becomes shorter than the age of the remnant, the shock wave starts to slow. When the interior pressure is exhausted the remnant then enters the snowplough phase with $r_{\rm shell} \propto t^{1/4}$ and is driven by momentum (momentum is conserved). 

The evolution of the shell radius with time is shown in Figure~\ref{fig:shellrad}. The shell is defined by taking the maximum density averaged radially in bins of 1pc. The dense shell starts to form at ~0.04Myr. As the Figure shows, after this time, the shell expands according to the Sedov-Taylor expectation until at least $1$\,Myr.  The initial evolution is due to the fact that we have injected thermal energy only and it therefore takes some time for this energy to be converted into radial momentum and for a well-defined shell to form.
Similarly, the evolution of the radial velocity of the shell with time is shown in Figure~\ref{fig:shellvel}. Here the shell is defined by taking the average radial velocity in radial bins of width 1pc; the particles in the velocity peak are at slightly larger radii than those with in the density peak. Initially, the velocity is low as the energy injected is thermal; this is gradually converted to kinetic energy and after $\sim6\times10^{4}$ years the resulting radial velocity is within a factor of two of the velocity predicted by the Sedov-Taylor solution. 
 
Figure~\ref{fig:boxmom} shows the evolution of total momentum (blue) and the radial component of momentum (red) in boxes of size 33pc and 50pc around the SNa. We chose the radius 33pc for the smaller box as this corresponds to our estimated radius of the shell at approximately $0.5$\,Myr. As the plot clearly shows, all the momentum in the gas is radial and therefore due to the energy input from the SNa event - the drop in radial momentum within 33pc at around $1$Myr corresponds to the time when the majority of the shell leaves the 33pc box but is still included in the 50pc box. The growth of the radial momentum follows the Sedov solution, albeit with an amplitude that is about a factor of two lower. We therefore conclude that at the resolution of our simulations, the representation of SNa explosions by thermal energy injections of $\sim 10^{50}$\,ergs results in approximately the correct amount of radial momentum being injected into the gas on a length scale of tens of parsecs. As the radial momentum injected is the key factor in determining the evolution of the ISM, our simulations capture all the essential physics of SNa feedback.

\endgroup


\bsp	
\label{lastpage}
\end{document}